\documentclass[fleqn,10pt]{wlscirep}
\usepackage[utf8]{inputenc}
\usepackage[T1]{fontenc}
\title{Precise phase retrieval for propagation-based images using discrete mathematics}

\usepackage{cite}
\usepackage{amsmath,amssymb,amsfonts}
\usepackage{algorithmic}
\usepackage{graphicx}
\usepackage{epstopdf}
\usepackage{bm}% bold math
\usepackage{siunitx}
\usepackage{textcomp}
\usepackage{subcaption}  
\usepackage{graphicx,caption}
\usepackage{tikz}
\usepackage{siunitx}

\author[1,*]{James A. Pollock}
\author[1]{Kaye S. Morgan}
\author[1]{Linda C.P. Croton}
\author[1]{Michelle K. Croughan}
\author[1]{Gary Ruben}
\author[2]{Naoto Yagi}
\author[2]{Hiroshi Sekiguchi}
\author[1]{Marcus J. Kitchen}
\affil[1]{School of Physics \& Astronomy, Monash University, VIC 3800, Australia (e-mail: james.pollock@monash.edu).}
\affil[2]{Japan Synchrotron Radiation Research Institute (JASRI)/SPring-8, Sayo, Hyogo, 679-5198, Japan}

\affil[*]{James.pollock@monash.edu}

\begin{abstract}
The ill-posed problem of phase retrieval in optics, using one or more intensity measurements, has a multitude of applications using electromagnetic or matter waves. Many phase retrieval algorithms are computed on pixel arrays using discrete Fourier transforms due to their high computational efficiency. However, the mathematics underpinning these algorithms is typically formulated using continuous mathematics, which can result in a loss in spatial resolution in the reconstructed images. Herein we investigate how phase retrieval algorithms for propagation-based phase-contrast X-ray imaging can be rederived using discrete mathematics and result in more precise retrieval for single- and multi-material objects and for spectral image decomposition. We validate this theory through experimental measurements of spatial resolution using computed tomography (CT) reconstructions of plastic phantoms and biological tissue, using detectors with a range of imaging system point spread functions (PSFs). We demonstrate that if the PSF substantially suppresses high spatial frequencies, the potential improvement from utilising the discrete derivation is limited. However, with detectors characterised by a single pixel PSF (e.g. direct, photon-counting X-ray detectors), a significant improvement in spatial resolution can be obtained, demonstrated here at up to 17\%.
\end{abstract}
\begin{document}

\flushbottom
\maketitle
\thispagestyle{empty}

\section{Introduction}
\label{sec:introduction}
X-ray imaging, through non-invasive single projection and Computed Tomography (CT), can provide excellent structural detail of samples and is a crucial tool in medical diagnosis, but it exposes the sample to harmful ionising radiation in the process. Although significant steps have been made to reduce the radiation dose, such as using iterative reconstruction methods \cite{kroft_added_2019}, spiral scans, and selective slice scans \cite{ball_ultra-low-dose_2017}, the incorporation of phase contrast (PC) techniques can be used to significantly reduce dose even further\cite{kitchen_ct_2017}. While several phase contrast techniques exist, we focus here on propagation-based phase-contrast imaging (PBI), taking advantage of the simplicity in optical design, which only requires a sufficiently coherent beam at the sample position and some propagation distance between the object and detector. Phase contrast appears as intensity fringes on the detector and arises from the interference of X-rays that have incurred different phase shifts after passing through different materials. This effect enables PBI to be used effectively to enhance contrast between low-Z materials. To reconstruct the sample properties from the images requires application of an appropriate phase retrieval algorithm. Since we cannot directly measure the phase of an X-ray wavefield, it must be derived from intensity images alone. Phase retrieval is therefore an ill-posed problem and few algorithms provide an accurate and robust solution. The phase retrieval algorithm of Paganin \textit{et al.} \cite{paganin_simultaneous_2002} provides such a solution for the case of homogenous objects, that is, those comprised of a single material, and converts the phase contrast effects to a signal-to-noise-ratio (SNR) boosted version of the absorption contrast image. This enables low-dose distinction between low-Z materials \cite{beltran_interface-specific_2011,kitchen_ct_2017}. The boost in SNR comes from the phase retrieval applying a customised blurring filter to the images, returning the phase-contrast sharpened boundaries to more directly match the object shape while removing high frequency noise in the process \cite{gureyev_unreasonable_2017}. The most common algorithm for this is a Lorentzian Fourier filter, based on the transport of intensity equation (TIE), which is referred to herein as the Paganin Method (PM). 

The PM relies on the paraxial approximation, requires a homogeneous object, and a small  propagation distance between the sample and detector $\Delta$ so that the resulting image is in the near-field regime. While material inhomogeneity leads to erroneous thickness determinations in projection-based images, the single-material requirement was later loosened by Beltran \textit{et al.} \cite{beltran_2d_2010} to allow for dual-material samples. See also a simplification when combined with CT in Croton \textit{et al.} \cite{croton_situ_2018}, which allows phase retrieval to be performed on interfaces between any two known materials, regardless of how many other materials are present in the sample. In its original form, the Paganin method of phase retrieval reconstructs the contact intensity as
\begin{align}
I_{\text{PM}}(x, y) &= \mathcal{F}^{-1}\left\{\frac{\mathcal{F}[I(x, y, z = \Delta)/I_0]}{1 + \frac{\delta \Delta}{\mu}\textbf{k}_{\perp \text{PM}}^2}\right\} \label{equation: paganin method thickness},
\end{align}
where $I_0$ is the incident wavefield intensity, $\delta$ is the decrement component of the complex refractive index $n = 1 - \delta + i \beta$, and $\mu$ is the linear attenuation coefficient, which is related to the imaginary component of the complex refractive index $\beta$ and x-ray wavelength $\lambda$ by $\mu= 4\pi\beta/ \lambda$. The vector $\textbf{k}_{\perp \text{PM}}^2$ represents the perpendicular components of the wavevector, defined as 
\begin{align}
\textbf{k}_{\perp \text{PM}}^2 = k_{x}^2 + k_{y}^2,\label{equation: PM kperp}
\end{align}
with ($k_{x}$,$k_{y}$) the Fourier spatial frequencies corresponding to the image coordinates $(x,y)$. 

In conjunction with CT,equation \eqref{equation: paganin method thickness} has been applied in a range of settings, including investigations of self-healing thermoplastics \cite{mookhoek_applying_2010}, studies of sandstone \cite{yang_data-constrained_2013} and coal micro-structure \cite{wang_evaluation_2015} as well as in animal imaging studies such as detecting iron oxide nanoparticles in mouse
brains \cite{marinescu_synchrotron_2013}\cite{rositi_information-based_2013}, and dose optimization of lung microtomography \cite{lovric_dose_2013}\cite{kitchen_ct_2017}. Of particular interest are the studies that report improvements to spatial resolution after incorporating sharpening filters to the algorithm, effectively suppressing high spatial frequencies to a lessor degree than implemented by the Fourier Lorentzian filter in equation \eqref{equation: paganin method thickness}. These studies included the ANKAphase \cite{weitkamp_ankaphase_2011} version 2.1 software package implementation of the PM incorporating a deconvolution filter, and the addition of an unsharp mask to the pyHST2 implementation \cite{mirone_pyhst2_2013}. Alternative methods have also included adding high spatial frequency information from the phase contrast image back into the retrieved images \cite{irvine_simple_2014}, intended as a compromise between phase retrieval and phase contrast. This motivated Paganin et al. \cite{paganin_boosting_2020} to revisit the derivation and find a first principles justification for the success of these approaches \cite{paganin_boosting_2020}. Previous, first-principles-supported methods for increasing the spatial resolution of the PM have broadened the algorithm's scope, such as by reducing the filter strength to account for inherent blurring by the system point spread function (PSF) \cite{beltran_phase-and-amplitude_2018}. However, with an algorithm based on fundamental wave optics, namely the new Generalised Paganin Method (GPM) in \cite{paganin_boosting_2020}, we can expect a correction that more accurately restores high spatial frequencies than post-hoc sharpening filters. 

The over-suppression of high spatial frequencies by the PM is a consequence of applying equation \eqref{equation: paganin method thickness}, derived using continuous Fourier transform integrals, to discrete pixel-based imaging systems. Note that, although equation \eqref{equation: paganin method thickness} includes Fourier transforms, this algorithm is implemented using discrete Fourier transforms, since digital images have discrete sampling. In the GPM derivation presented by Paganin \textit{et al.} \cite{paganin_boosting_2020}, this is addressed by employing a 5-point approximation of the transverse Laplacian \cite{press_numerical_1992}\cite{abramowitz_handbook_1965} into which the discrete representation of the Fourier transform is substituted. This representation is then used, in place of the Fourier derivative theorem, to expand the Laplacian that appears in the derivation of the original PM. The contact plane intensity in the GPM is given as
\begin{align}
I_{\text{GPM}}(x, y) &= \text{DFT}^{-1}\left[\frac{\text{DFT}[I(x, y, z = \Delta)/I_0]}{1 + \frac{\delta \Delta}{\mu}\textbf{k}_{\perp \text{GPM}}^2}\right], \label{equation: general paganin method thickness}
\end{align}
where $\text{DFT}$ represents the discrete Fourier transform mapping from the image coordinates $(x, y)$ to the Fourier space coordinates $(k_{x},k_{y})$, and $\text{DFT}^{-1}$ is the inverse transform. Equation \eqref{equation: general paganin method thickness} explicitly accounts for the discrete image sampling at coordinates $(x,y)$, while possessing a similar form as the PM in equation \eqref{equation: paganin method thickness} differing only in the quantification of the spatial frequencies, now represented by  
\begin{align}
\textbf{k}_{\perp \text{GPM}}^2 = \frac{-2}{W^2}[\cos(W k_{x}) + \cos(W k_{y}) - 2], \label{equation: GPM kperp}
\end{align}
where $W$ is the pixel size of the detector. As demonstrated in Paganin $et al.$ \cite{paganin_boosting_2020}, a Taylor series expansion of equation \eqref{equation: GPM kperp} shows convergence to equation \eqref{equation: PM kperp} for spatial frequencies close to the origin of Fourier space, but this can differ greatly when approaching the Nyquist frequency of the Fourier transform. Paganin \textit{et al.} \cite{paganin_boosting_2020} demonstrated that the GPM filter always suppresses high spatial frequencies to a lesser extent than the PM, providing some first principles justification for the spatial resolution improvement found by users applying sharpening masks alongside the PM algorithm\cite{paganin_boosting_2020}. Paganin \textit{et al.} \cite{paganin_boosting_2020} provided a comparison between the PM and GPM spatial frequency filters by varying $\delta \Delta / \mu W$ over several orders of magnitude. They also demonstrate a difference between images reconstructed with the GPM and PM algorithms by subtracting one from the other, but did not consider how the PSF may affect the utility of the GPM. Our study aims to quantify the improvement in spatial resolution of the GPM by looking at the PSF of the entire imaging and reconstruction system, and identifying the experimental conditions under which the GPM provides noticeably improved spatial resolution compared to the PM. 

We begin in Section \ref{section: spatial frequency filter} by analyzing the respective transfer functions of each algorithm, comparing their spatial frequency filters before incorporating the system PSF to describe how well the imaging and retrieval system captures the various spatial frequencies present in the sample. Section \ref{section: circle simulations} then provides a brief comparison of the resolution each can achieve on a simulated phantom projection. Section \ref{sec: indirect detectors} explores improvements achieved in CT and compares the differences between direct and indirect X-ray imaging detectors. Having determined the best experimental configuration to exploit the utility of the GPM, we demonstrate in Section \ref{section: established algorithms} that the discrete Fourier transform notation introduced in the GPM can also be used to prevent over-blurring in other phase retrieval algorithms that use DFTs. Specifically, we experimentally demonstrate more accurate image reconstruction, via improved spatial resolution, for the two-material phase retrieval algorithm of Beltran \textit{et al.} \cite{beltran_2d_2010} (Section \ref{section: established algorithms - beltran}) and spectral decomposition algorithm of Schaff \textit{et al.} \cite{schaff_spectral_2020} (Section \ref{section: established algorithms - Florian}).

\section{Simulated Analysis in Projection}
\label{section: Projection}
We begin with preliminary comparisons of the PM and GPM methods by looking at the respective spatial filters. We incorporate additional filtering to mimic other stages of imaging, to better emulate how spatial frequencies in the sample are captured in the raw image and then appear in the final retrieved sample image. These two methods are also applied to simulated projection images to measure the resulting spatial resolution of the imaging system. 

\subsection{Analysis of system transfer functions}
\label{section: spatial frequency filter}
\begin{figure*}[b!]
	\begin{tikzpicture}
	\node[anchor=south west] at (0,0) (image1) {\includegraphics[width=.48\textwidth]{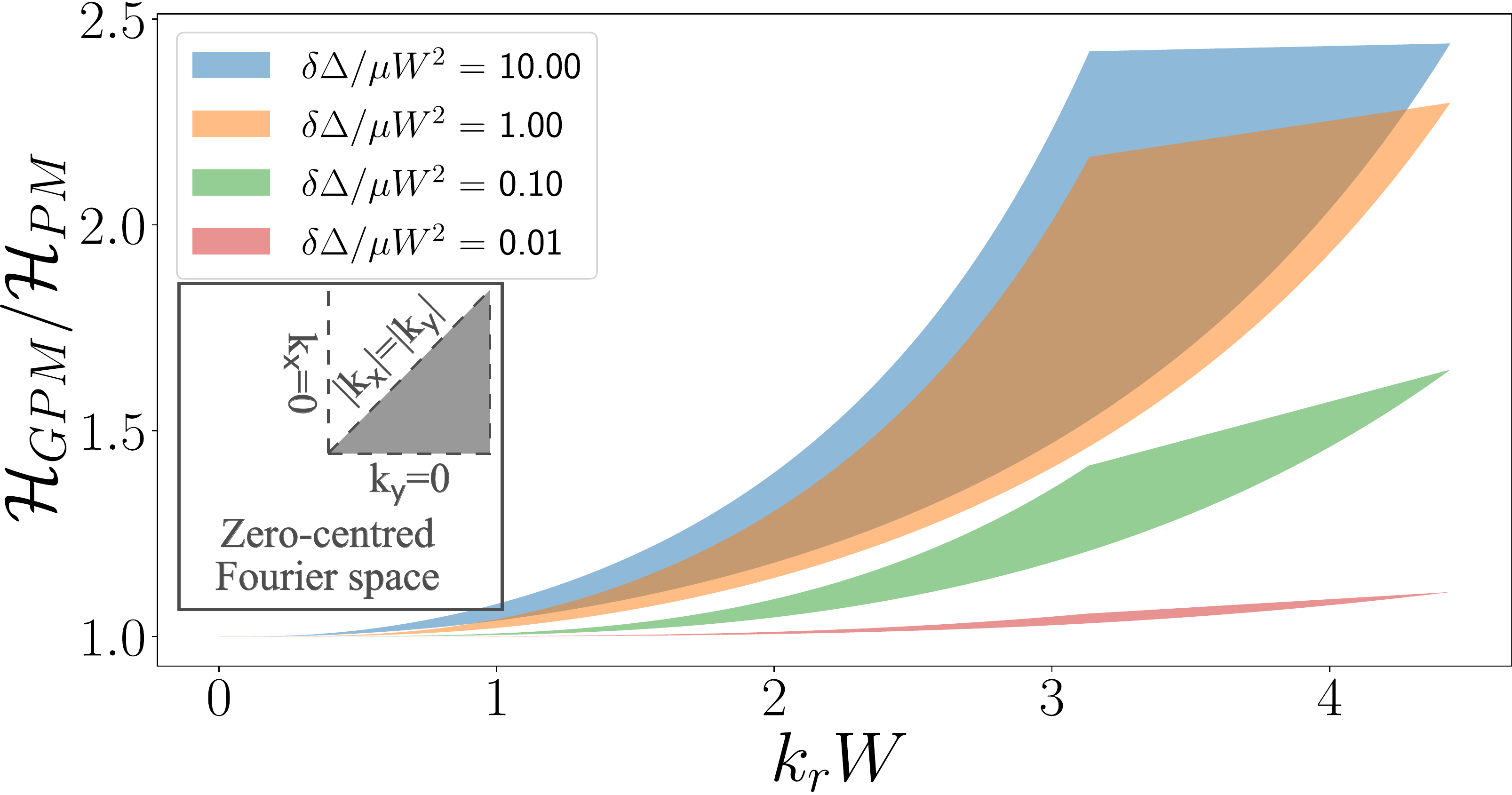}};
	% draw a grid and draw ticks
	\begin{scope}[x={(image1.south east)},y={(image1.north west)}]
	%\draw[help lines,xstep=.1,ystep=.1,overlay] (0,0) grid (1,1);
	%\foreach \x in {0,1,...,9} { \node [anchor=north,overlay] at (\x/10,0) {0.\x}; }
	%\foreach \y in {0,1,...,9} { \node [anchor=east,overlay] at (0,\y/10) {0.\y}; }
	\node[fill=none, font = \large] at (0.03,0.9) {a)};
	\end{scope}
	\end{tikzpicture}\hfill
	\begin{tikzpicture}
	\node[anchor=south west] at (0,0) (image1) {\includegraphics[width=.48\textwidth]{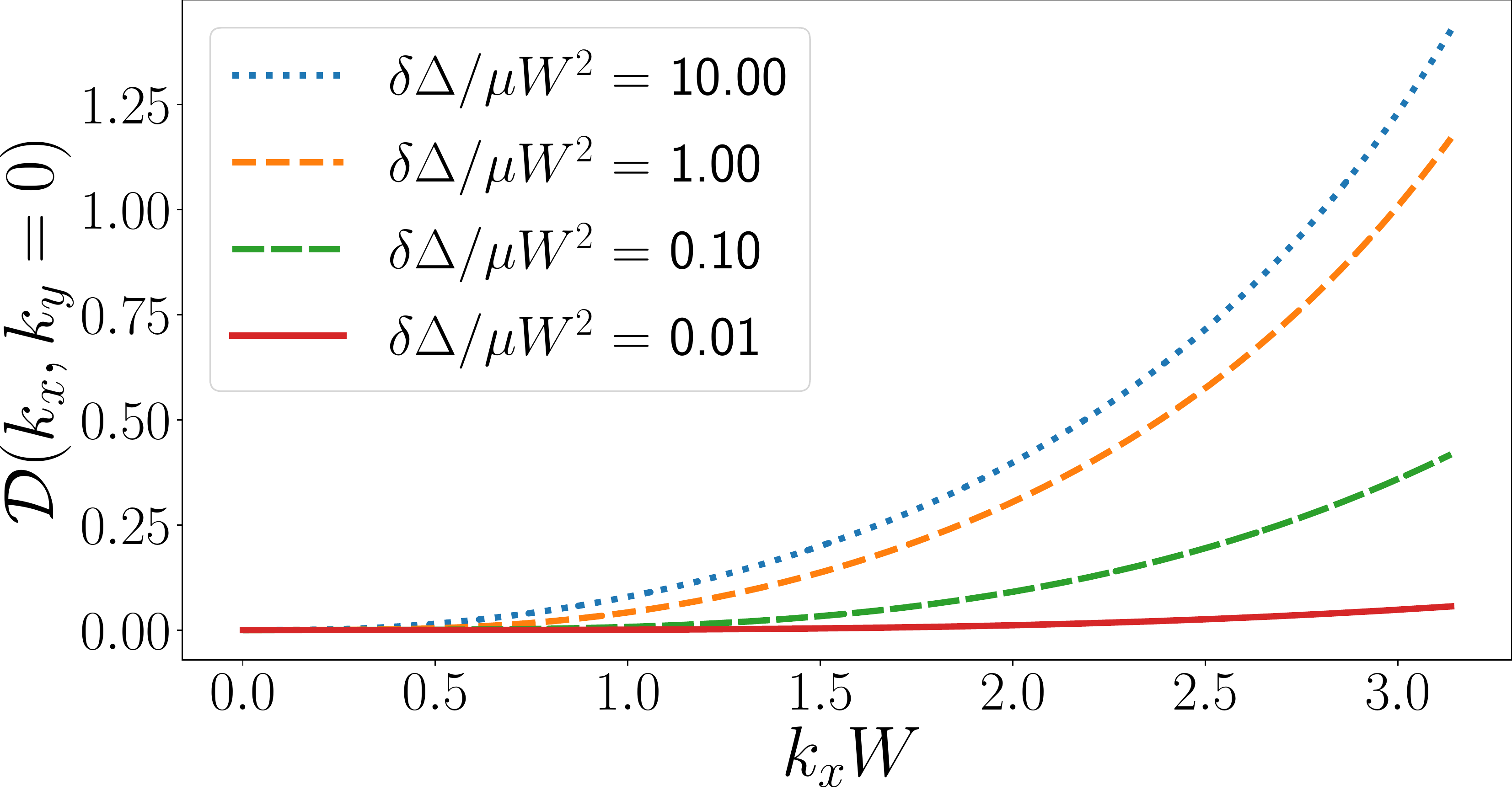}};
	% put the subfigure label in 
	\begin{scope}[x={(image1.south east)},y={(image1.north west)}]
	\node[fill=none, font = \large] at (0.03,0.9) {b)};
	\end{scope}
	\end{tikzpicture}\\%
	\begin{tikzpicture}
	\node[anchor=south west] at (0,0) (image1) {\includegraphics[width=.48\textwidth]{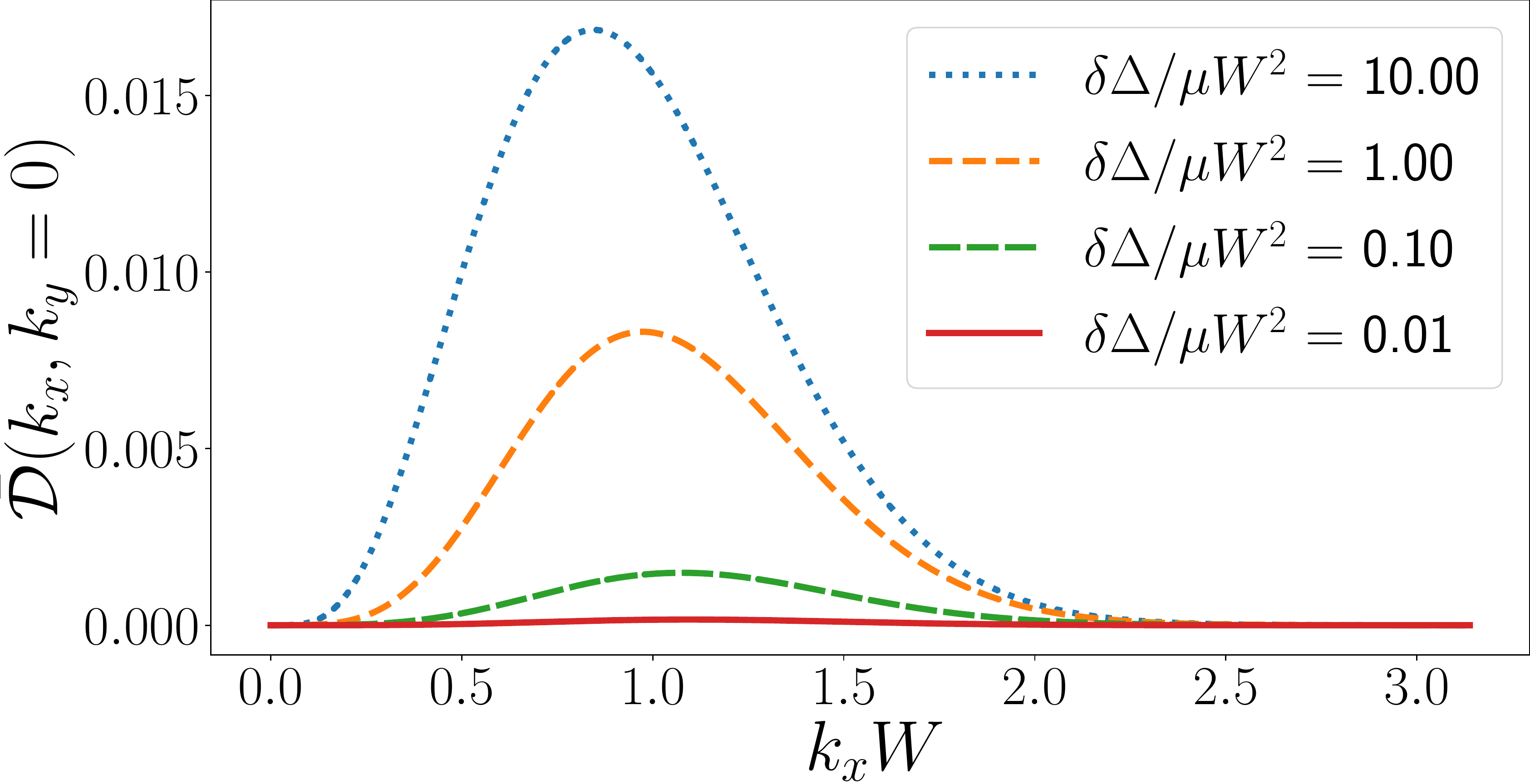}};
	% put the subfigure label in 
	\begin{scope}[x={(image1.south east)},y={(image1.north west)}]
	\node[fill=none, font = \large] at (0.03,0.9) {c)};
	\end{scope}
	\end{tikzpicture}
	\hfill
	\begin{tikzpicture}
	\node[anchor=south west] at (0,0) (image1) {\includegraphics[width=.48\textwidth]{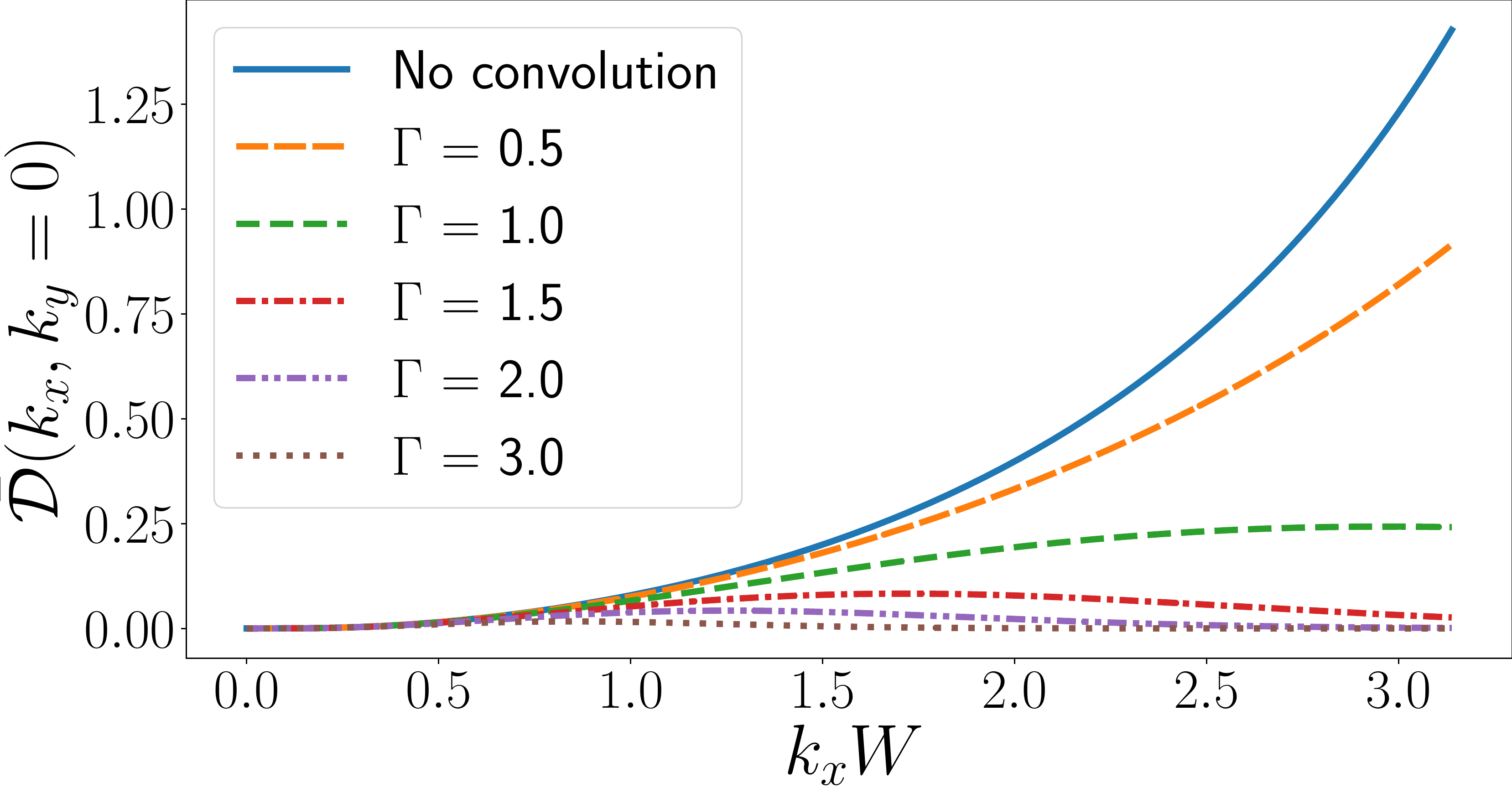}};
	% put the subfigure label in 
	\begin{scope}[x={(image1.south east)},y={(image1.north west)}]
	\node[fill=none, font = \large] at (0.03,0.9) {d)};
	\end{scope}
	\captionsetup{font = small}
	\end{tikzpicture}%
	\captionsetup{font = small}
	\caption{Comparative analysis of the PM and GPM transfer functions. (a) Displays the ratio of the phase retrieval transfer functions, equations \eqref{eq: GPM transfer function} and \eqref{eq: PM transfer function}, for various values of $\delta\Delta/\mu W^2$, using a horizontal and diagonal slice of the 2D filter to create a bounded region presenting the asymmetry of the GPM filter. (b) Plots the horizontal, ($k_x, k_y = 0$), line of the fractional difference in transfer functions described by equation \eqref{eq: fractional difference}, used as a comparison to (c) the imaging system transfer function, equation \eqref{eq: combined fractional difference}, which incorporates a $\mathcal{G}(k_{x}, k_{y}, \Gamma)$ PSF, set as $\Gamma=3.0$ FWHM which reduces the plot's vertical scaling, as well as the phase retrieval transfer function. Finally, (d) directly displays the effect of varying the PSF width, $\Gamma$, on the imaging system transfer function, for the case $\delta\Delta/\mu W^2 = 10$.  }
	\label{figure: spatial frequency suppressions}
\end{figure*}
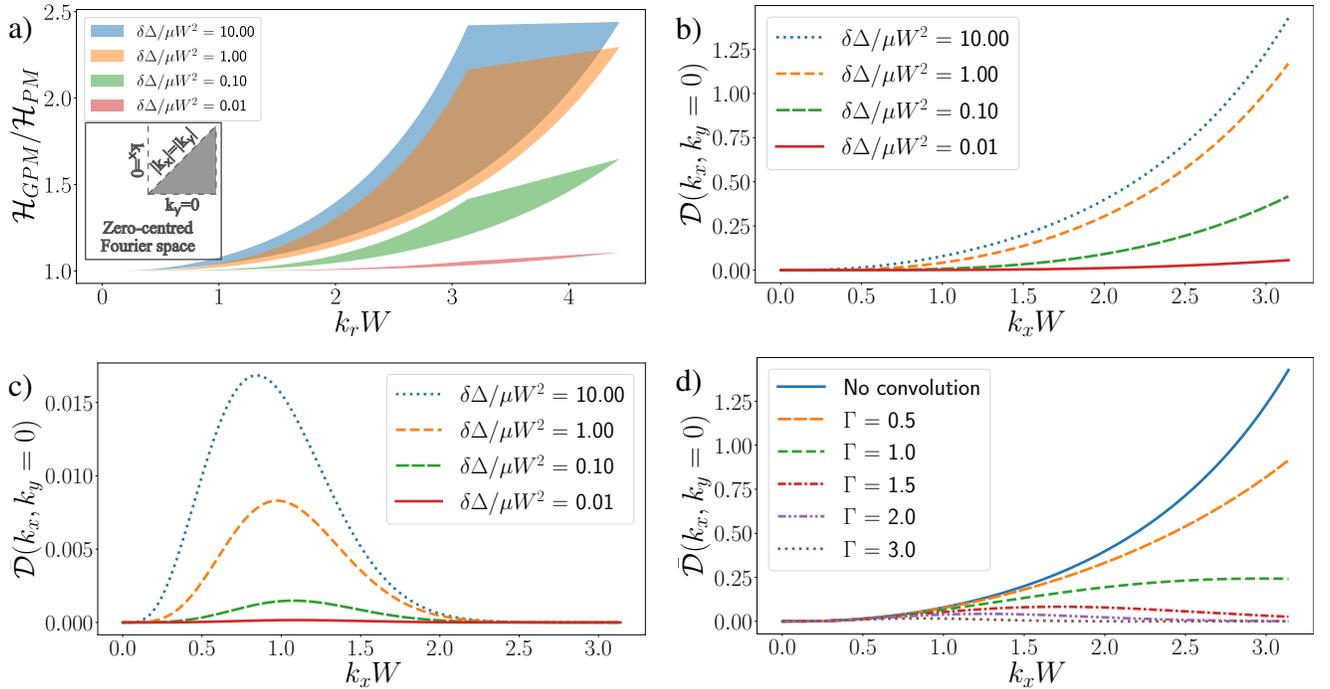%
The PM, equation \eqref{equation: paganin method thickness}, is ultimately a tailored spatial frequency filter, derived under the transport of intensity equation, used to blur an image such that the phase contrast seen at material boundaries is spread to reconstruct the sample thickness. Given that the only difference between the PM and GPM methods is the shape of this spatial frequency filter, this becomes our first point of comparison. From equations \eqref{equation: paganin method thickness} and \eqref{equation: general paganin method thickness}, we can define transfer functions for the application of each filter to raw images as 
\begin{align}
\mathcal{H}_\text{PM}(k_{x}, k_{y}) &= \frac{1}{1 + \frac{\delta \Delta}{\mu}(k_{x}^2 + k_{y}^2)}, \label{eq: PM transfer function}\\
\mathcal{H}_\text{GPM}(k_{x}, k_{y}) &= \frac{1}{1 - \frac{2 \delta \Delta}{\mu W^2}[\cos(W k_{x}) + \cos(W k_{y}) - 2]}, \label{eq: GPM transfer function}
\end{align}
given as Eqs. (20) and (19) in Paganin \textit{et al.} \cite{paganin_boosting_2020}, where $\mathcal{H}_\text{PM}(k_{x}, k_{y})$ represents the amplification applied to each spatial frequency amplitude by the PM, and $\mathcal{H}_\text{GPM}(k_{x}, k_{y})$ the amplification applied by the GPM. From here, a simple comparison between equations \eqref{eq: PM transfer function} and \eqref{eq: GPM transfer function} can be performed by taking their ratio (equation (21) in [2]), $R(k_x,k_y) = \mathcal{H}_\text{GPM}(k_{x}, k_{y})/\mathcal{H}_\text{PM}(k_{x}, k_{y})$ and plotting the result on a normalized axis. Figure \ref{figure: spatial frequency suppressions}(a) does this for a range of values of the combined parameter $\delta\Delta/\mu W^2$, displaying the fractional difference between the GPM and PM as a shaded region bounded by the $|k_y| \cup |k_x|= 0$ (upper bound) and $|k_x| = |k_y|$ (lower bound) lines. For the $|k_x| = |k_y|$ lines, seen as the lower bound of the Fig. \ref{figure: spatial frequency suppressions}(a) plots, we produce the one-dimensional spatial frequency axis through $k_r = W\sqrt{k_{x}^2 + k_{y}^2}$, leading to values above $\pi$ where $(k_x, k_y)$ extend into the corners of a square image. The shaded regions help to demonstrate the new-found asymmetry of the GPM filter, correcting the PM algorithm to deliver a more uniform treatment of all edges in real-space, regardless of their orientation. A more helpful way to reveal quantitative differences between each algorithm is through the fractional difference, defined as 
\begin{align}
\mathcal{D}(k_{x}, k_{y}) &= \frac{\mathcal{H}_\text{GPM}(k_{x}, k_{y}) - \mathcal{H}_\text{PM}(k_{x}, k_{y})}{\mathcal{H}_\text{PM}(k_{x}, k_{y})}, \label{eq: fractional difference}
\end{align}
which converges toward zero when the PM and GPM filters match, as opposed to the ratio of equations \eqref{eq: PM transfer function} and \eqref{eq: GPM transfer function} which converges to one. Figure \ref{figure: spatial frequency suppressions}(b) plots equation \eqref{eq: fractional difference} across the $k_x$ or $k_y$ axis line ($|k_y| \cup |k_x| = 0$). We see again that the larger frequencies experience the greatest difference between the two algorithms, leading to increased spatial resolution in the GPM due to the greater proportion of high spatial frequencies, and that they are effectively indistinguishable at the lower spatial frequencies. However, equations \eqref{eq: PM transfer function}-\eqref{eq: fractional difference} only represent the transfer function of the post-image processing, whereas to better reflect experimental conditions we must also account for the blurring effect of the detector imaging system and optical system (i.e. all blurring effects aside from the propagation and phase retrieval). To do this we introduce the contrast transfer function $\mathcal{G}(k_{x}, k_{y}, \Gamma)$. We describe the real-space detector PSF as an azimuthally symmetric Gaussian, so that $\mathcal{G}(k_{x}, k_{y}, \Gamma)$ is given by the Fourier transform 
\begin{align}
\mathcal{G}(k_{x}, k_{y}, \Gamma) &= \mathcal{F}\left[\exp(-\frac{x^2+y^2}{\Gamma^2/2.355^2})\right]\\ &= \exp(-\Gamma^2(4\pi^2k_{x}^2 + 4\pi^2k_{y}^2)/2.355^2), \label{eq: gaussian FT}
\end{align}
where $\Gamma$ is the a full width at half maximum (FWHM) in real space, measured in pixels, and the factors of $4\pi^2$ in the exponential reflect the DFT normalization convention. Combining the imaging system transfer function, equation \eqref{eq: gaussian FT}, with the phase retrieval transfer functions, equations \eqref{eq: PM transfer function}-\eqref{eq: GPM transfer function}, gives the complete transfer functions $\bar{\mathcal{H}}$ as
\begin{align}
\bar{\mathcal{H}}_\text{PM}(k_{x}, k_{y}) &= \mathcal{H}_\text{PM}(k_{x}, k_{y})\mathcal{G}(k_{x}, k_{y}, \Gamma),\label{eq: combined PM transfer function}\\
\bar{\mathcal{H}}_\text{GPM}(k_{x}, k_{y}) &= \mathcal{H}_\text{GPM}(k_{x}, k_{y})\mathcal{G}(k_{x}, k_{y}, \Gamma),\label{eq: combined GPM transfer function}
\end{align}
allowing us to similarly define a new fractional difference as
\begin{align}
\bar{\mathcal{D}}(k_{x}, k_{y}) &= \frac{\bar{\mathcal{H}}_\text{GPM}(k_{x}, k_{y}) - \bar{\mathcal{H}}_\text{PM}(k_{x}, k_{y})}{\mathcal{H}_\text{PM}(k_{x}, k_{y})}. \label{eq: combined fractional difference}
\end{align}
By incorporating the blurring effects of the imaging and imaging system into \ref{eq: fractional difference}, we can expand on the filter comparisons in Paganin \textit{et al.} \cite{paganin_boosting_2020} and demonstrate how imaging PSFs may limit the relative spatial resolution improvement between the algorithms. Figure \ref{figure: spatial frequency suppressions}(c) plots equation \eqref{eq: combined fractional difference} for the same $\delta\Delta/\mu W^2$ values as in Fig. \ref{figure: spatial frequency suppressions}(b), now including a Gaussian PSF with FWHM of 3 pixels. We see that the amplitudes of all spatial frequencies are reduced relative to panels (a) and (b), particularly at the higher spatial frequencies, and the difference between the GPM and PM is reduced overall. This predicts that there may only be a very small difference in spatial resolution between the PM and GPM methods when the PSF $\Gamma$ is a few pixels wide, as is typical of most indirect X-ray detectors. We also see that the biggest difference is now shifted to medium spatial frequencies in this example. Figure \ref{figure: spatial frequency suppressions}(d) plots equation \eqref{eq: combined fractional difference} for $\delta\Delta/\mu W^2=10$, while varying the PSF size in pixels ($\Gamma$). The `no convolution' trend displays the fractional difference without incorporating a PSF, $\mathcal{D}(k_{x}, k_{y})$, as a reference point. We observe that increasing the PSF size decreases the amplitudes of high spatial frequencies, hence likely decreasing the potential improvement to resolution available via the GPM. However, for PSF widths around 1 pixel, typical of direct X-ray detectors, such as photon counting detectors, the high spatial frequency amplitudes are still 20\% increased under the GPM algorithm. This leads us to suspect that direct detectors, such as photon-counting detectors, will be best suited to benefit from the GPM algorithm, while for indirect X-ray detectors, which can possess PSF widths of two or more pixels, the improvement may be relatively minor.

\subsection{Spatial resolution improvement in projection images}
\label{section: circle simulations}
\begin{figure*}[t!]
	\begin{tikzpicture}
	\node[anchor=south west] at (0,0) (image1) {\includegraphics[width=.48\textwidth]{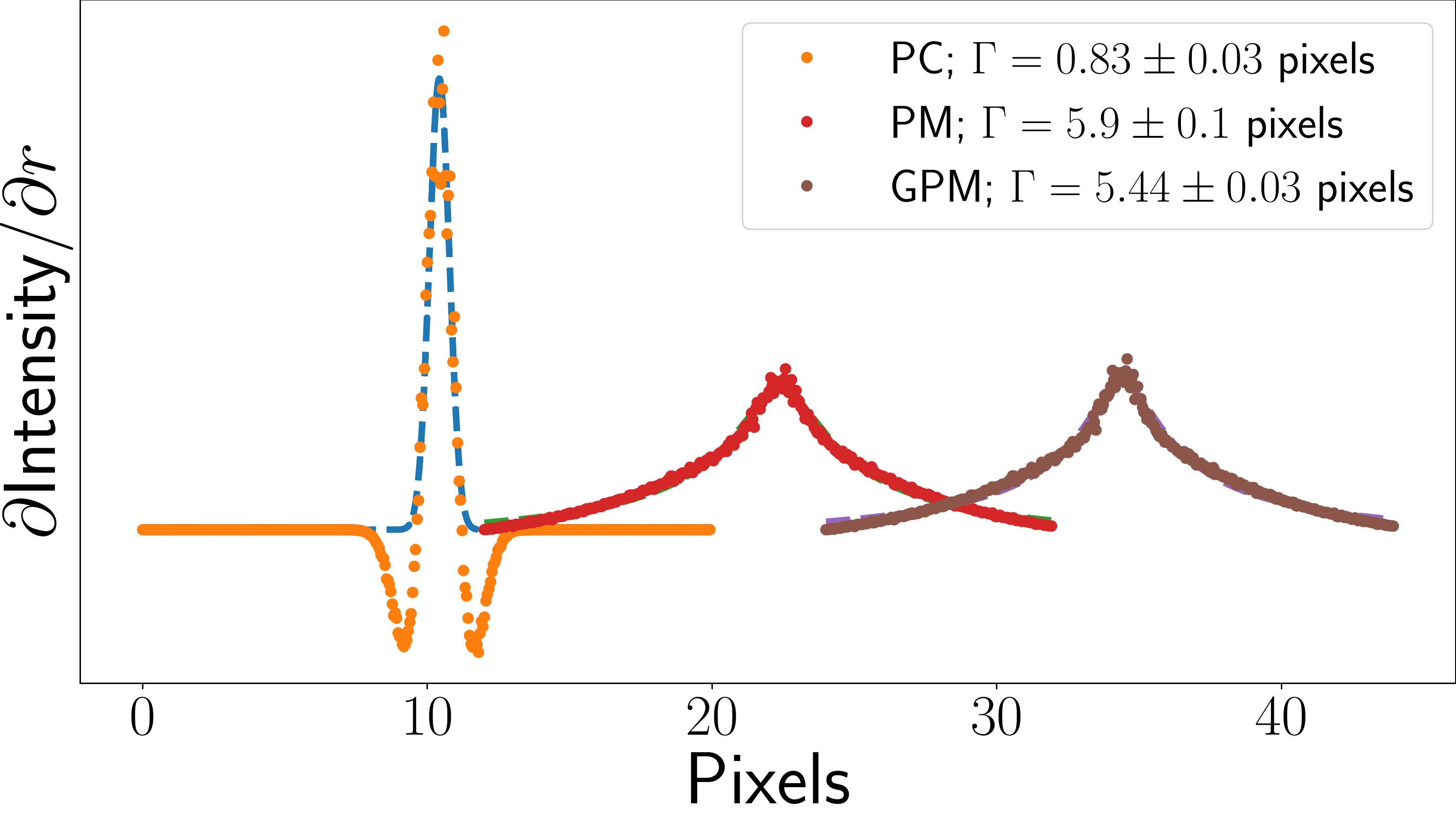}};
	% draw a grid and draw ticks
	\begin{scope}[x={(image1.south east)},y={(image1.north west)}]
		\node[fill=none, font = \large] at (0.03,0.9) {a)};
	\end{scope}
	\end{tikzpicture}\hfill
	\begin{tikzpicture}
	\node[anchor=south west] at (0,0) (image1) {\includegraphics[width=.48\textwidth]{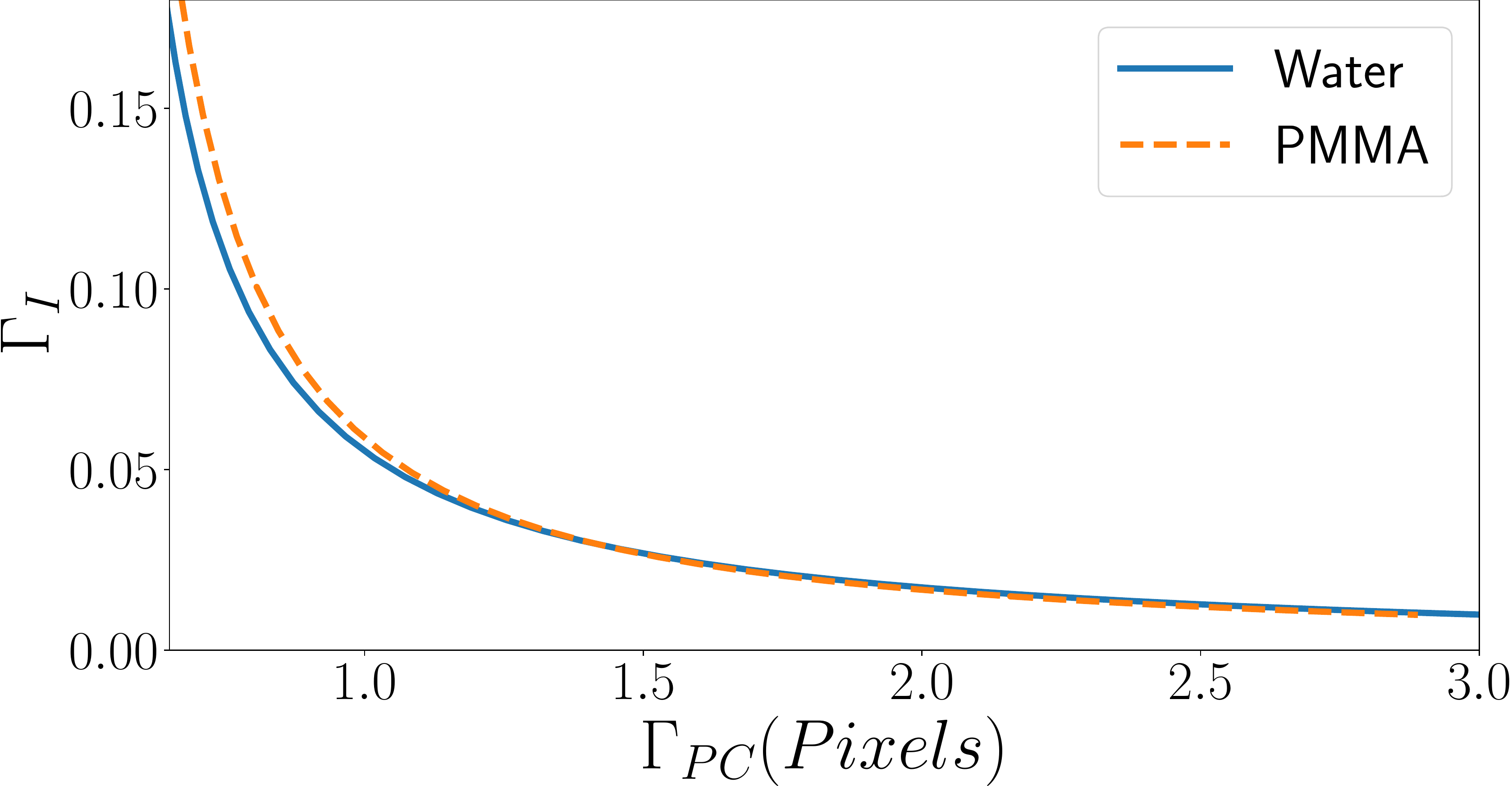}};
	% draw a grid and draw ticks
	\begin{scope}[x={(image1.south east)},y={(image1.north west)}]
		\node[fill=none, font = \large] at (0.03,0.9) {b)};
	\end{scope}
	\end{tikzpicture}
	\captionsetup{font = small}
	\caption{(a) Azimuthally averaged imaging system Line Spread Functions (LSF)s of the circular phantom image showing the effect of phase retrieval on spatial resolution from the phase contrast (PC) and phase retrieved images using the PM and GPM algorithms for a sample composed of water. Underlying dashed curves represent Pearson VII fits used to measure the LSF width. The phase contrast PSF was rescaled vertically by a factor of 8 for plotting. (b) Plots the percentage improvement in resolution, according to equation \eqref{equation: psf percentage improvement} of the GPM, plotted against the initial resolution of the simulated object. }
	\label{figure: simulation percentage improvement with PC PSF}
\end{figure*}%
The previous section showed how the shape of the image transfer function can vary between the PM and GPM phase retrieval algorithms as a result of adding realistic PSFs to the system. Here we quantify what effect the PSF has on spatial resolution in combination with phase retrieval via numerical simulation. We performed the comparison by simulating the propagation of a wavefield through cylindrical phantoms using the TIE (details below) until phase contrast fringes were produced. Next, we applied each phase retrieval algorithm and created an azimuthally-averaged profile of the phantom edge, which was differentiated to create a line spread function (LSF) that can be measured to evaluate the spatial resolution of the phase-retrieved image. We then use the measured resolution in the propagation-based phase contrast image, pre-phase retrieval, as a basis for our resolution comparison. Note that, while LSFs are one dimensional, and PSFs are two dimensional, both are distinct measurements of spatial resolution entities, and we will use the terms interchangeably throughout the paper. A typical LSF is well approximated by a Pearson VII function,
\begin{align}
P(x) &= A\left[1 + \frac{4(x - x_0)^2}{\Gamma^2}\left(2^{\frac{1}{m}} - 1\right)\right]^{-m},%\label{equation: pearson VII},
\end{align}
where $x_0$ is the peak position, $\Gamma$ is the FWHM and $m$ is the exponent that sets the position on the spectrum between Lorentzian ($m=1$) and Gaussian (approximated by $m>10$) behaviour. Finally, we use the FWHM values of the imaging system PSF measured in the phase-retrieved images from the PM ($\Gamma_{\text{PM}}$) and GPM ($\Gamma_{\text{GPM}}$) to calculate a fractional improvement in spatial resolution between the two algorithms,
\begin{align}
\Gamma_{\text{I}} &= \frac{\Gamma_{\text{PM}} - \Gamma_{\text{GPM}}}{\Gamma_{\text{PM}}}. \label{equation: psf percentage improvement}
\end{align}
Our simulations used end-on cylindrical phantoms composed of water (\SI{0.998}{\gram\per\cubic\cm}) and polymethyl methacrylate (PMMA, \SI{1.19}{\gram\per\cubic\cm}) with a radius of $900.5\:$pixels, created on a $2048\times2048$ pixel array with pixel size W = \SI{25}{\micro\meter}. The wavefield directly after transmission through the phantom was constructed using the projection approximation \cite{morgan_projection_2010} on a $\times5$ up-sampled grid, assuming an object thickness of \SI{6}{\mm}, and propagated with the TIE until a single phase contrast fringe became visible in the wavefield intensity (\SI{4}{\mm}). A small Gaussian blurring filter ($\Gamma = 1.0\:$pixel) was applied to the thickness map pre-propagation to suppress artefacts arising from the pixelated boundaries of the circular phantom, before the second blurring filter was applied post-propagation to simulate a detectors with varying PSFs. Phase retrieval was then performed using either the PM or GPM methods. Figure \ref{figure: simulation percentage improvement with PC PSF}(a) shows example imaging system PSF measurements, before and after phase retrieval, incorporating a $\mathcal{G}(k_{x}, k_{y}, \Gamma)$ component, simulated through a Gaussian blurring filter, applied after the TIE propagation \cite{zuo_transport_2020}. 
 
From Fig. \ref{figure: simulation percentage improvement with PC PSF}(b) we see, for these low-Z materials, a $\sim$\SI{6}{\percent} improvement in the resolution when the phase contrast PSF FHWM is equal to the pixel size. This benefit reduces with increasing detector PSF width. At 2 pixels wide, a $\sim$\SI{2}{\percent} improvement is seen, and only a $\sim$\SI{1}{\percent} improvement is seen at 3 pixels PSF FWHM. This reinforces that the benefit of the GPM method is heavily dependent on the detector PSF and will show the greatest improvement over the PM when the detector PSF width is equivalent to a single pixel or smaller.

\section{Experimental analysis in CT}
Phase retrieval is often combined with CT to provide three-dimensional separation of materials with very high signal-to-noise ratios due to its three-dimensional smoothing operation \cite{kitchen_ct_2017}. We recorded tomographic scans of various phantoms to compare the PM and GPM using direct and indirect pixelated detectors to experimentally verify the effects of different detector PSFs on the reconstructions. We used synchrotron radiation, which provides high coherence and a low divergence beam, which are ideal for phase contrast imaging. To reconstruct the CT data, we first performed the standard flat field and dark current correction followed by phase retrieval, then used parallel beam filtered back projection with a ramp filter.

\subsection{Indirect X-ray detectors}
\label{sec: indirect detectors}
\begin{figure}[t!]
	\centering
	\begin{tikzpicture}
	\node[anchor=south west] at (0,0) (image1) {\includegraphics[width=.45\textwidth]{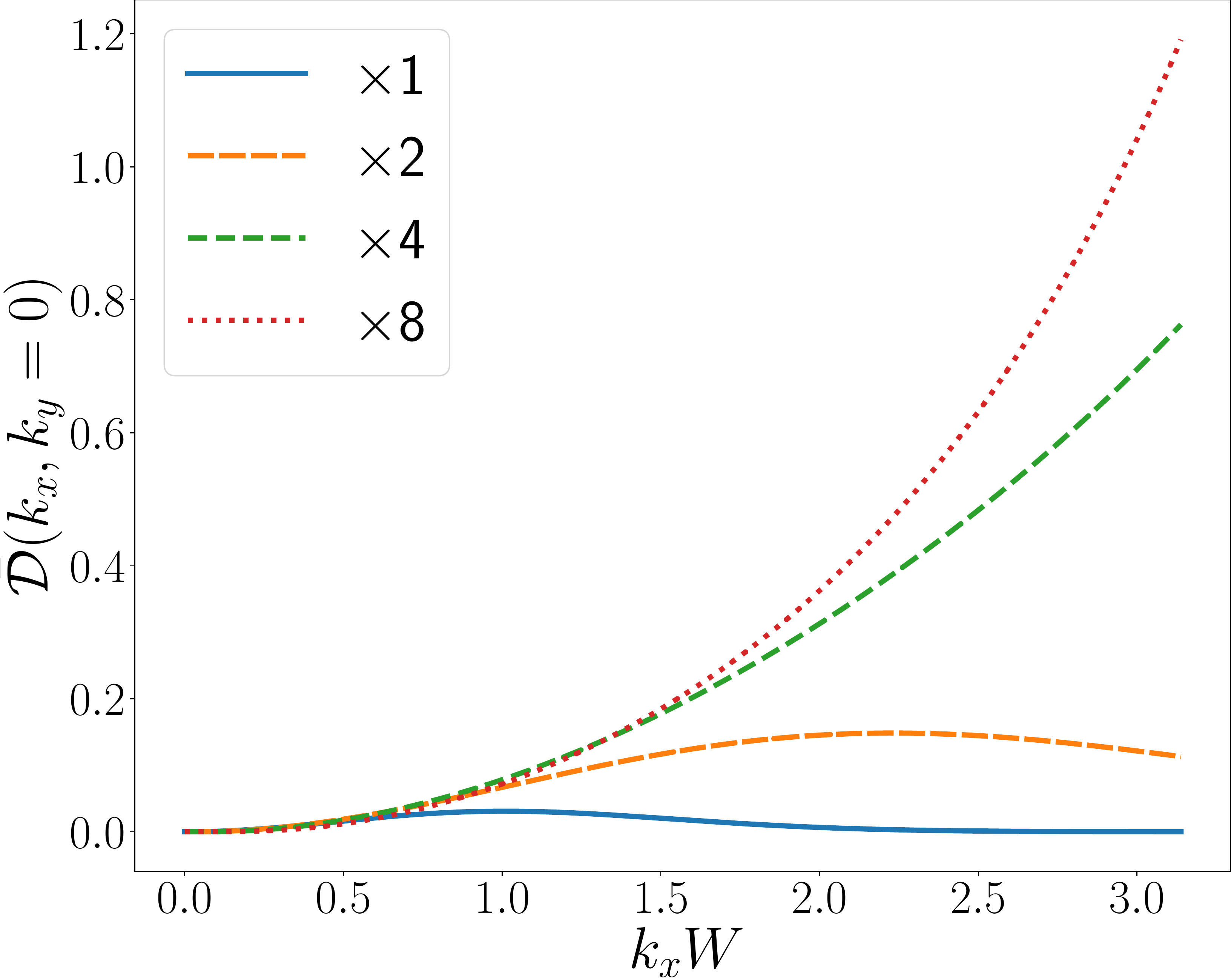}};
	% draw a grid and draw ticks
	\begin{scope}[x={(image1.south east)},y={(image1.north west)}]
		\node[fill=none, font = \large] at (0.03,0.9) {a)};
	\end{scope}
	\end{tikzpicture}%
	\begin{tikzpicture}
	\node[anchor=south west] at (0,0) (image1) {\includegraphics[width=.45\textwidth]{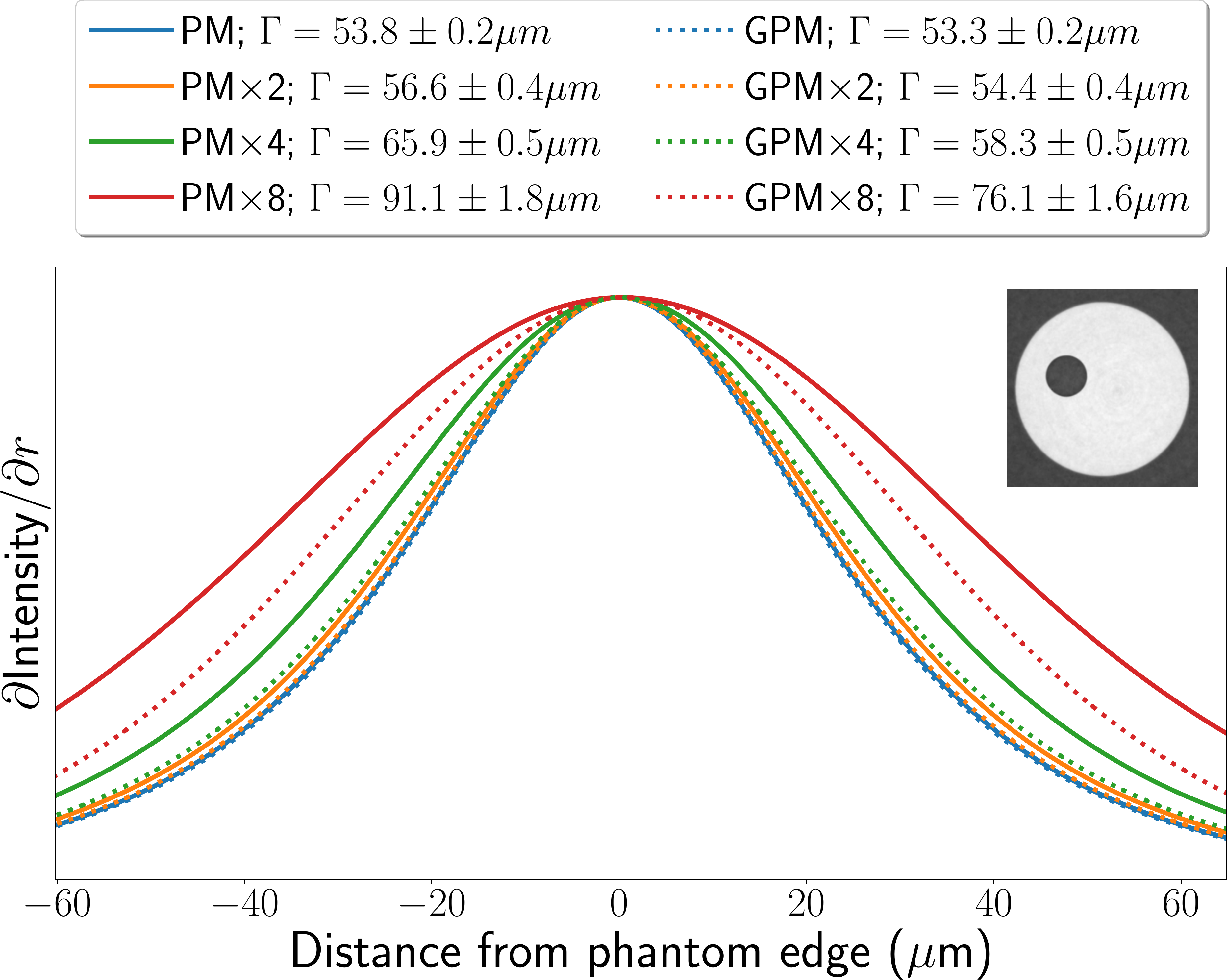}};
	% draw a grid and draw ticks
	\begin{scope}[x={(image1.south east)},y={(image1.north west)}]
		\node[fill=none, font = \large] at (0.03,0.9) {b)};
	\end{scope}
	\end{tikzpicture}
	\captionsetup{font = small}
	\caption{The effect of rebinning the raw phase contrast image ($\times n$) upon the spatial resolution of the phase-retrieved images, using the PM and GPM methods. (a) plots the transfer function fractional difference, equation \eqref{eq: combined fractional difference}, along the $(k_x,k_y=0)$ line, for various levels of rebinning. Here the filter conditions were chosen to initially match the detector PSF and the conditions used in	(b) which shows the Pearson VII fit reconstructions of imaging system PSF measurements conducted at the outer edge of a cylindrical PMMA phantom, shown in inset.}
	\label{figure: little flash rebinning}
\end{figure}%
Indirect X-ray detectors typically have a PSF of several pixels in width, introduced during the conversion of X-rays to visible light, where the divergent visible light from a single x-ray photon spreads across several pixels on the detector chip. Our investigation, which focused on indirect X-ray detectors, used a Hamamatsu Orca Flash 4.0 detector, with a \SI{10}{\micro\meter} thick Gadox (P43) phosphor directly coupled to the sCMOS sensor, with \SI{6.5}{\micro\meter} pixels in a $2048\times2048$ geometry. Combined with the imaging system at beamline 20B2 of the SPring-8 Synchrotron in Japan, with a propagation distance of $\Delta=$ \SI{2}{\meter} and energy of \SI{24}{\kilo\eV}, this produced a system PSF with FHWM $\Gamma=$ \SI{15.6}{\micro\meter} or $2.4\;$pixels. The filter shape analysis in Section \ref{section: Projection} suggests that such conditions would likely lead to negligible improvement in spatial resolution under the GPM. To reproduce this effect in experimental data, here we explore pixel rebinning as a method to probe different pixel sizes for a given detector PSF, $\mathcal{G}(k_{x}, k_{y}, \Gamma)$, and hence allow the GPM to demonstrate improvement as the pixel size approaches the PSF width. Rebinning is frequently used to increase the signal-to-noise ratio in circumstances where a longer exposure time is not desired, or to achieve fast data transfer times for high-speed imaging, so is important to consider in the context of this kind of experiment. Figure \ref{figure: little flash rebinning}(a) provides transfer function fractional differences, evaluated using equation \eqref{eq: combined fractional difference}, for the detector settings listed above, with increasing levels of rebinning. We observe that rebinning leads to larger differences between the PM and GPM at higher spatial frequencies, with the GPM potentially providing increased relative resolution at the new pixel size. Figure \ref{figure: little flash rebinning}(b) demonstrates the effect of rebinning on experimentally recorded CT slices created from 1800 projection images acquired over \SI{180}{\degree} rotation. As expected, analysis without rebinning produced only a negligible improvement in the width of the PSF after applying the GPM filter instead of the PM; approximately \SI{0.9\pm0.5}{\percent}. Rebinning by factors of 2, 4 and 8 along each axis show further improvements in the GPM spatial resolution compared to the PM. This improvement is quantified as a change in PSF width of \SI{4\pm1}{\percent} when rebinning by a factor of 2, \SI{12\pm2}{\percent} at a factor of 4, and \SI{16\pm2}{\percent} at a factor of 8. Each level of rebinning shows improvement; however, the difference between rebinning by 2 and 4 is far greater than that between rebinning by 4 and 8, as expected from the higher proportion of high spatial frequency information allowed by the transfer function (Fig. \ref{figure: little flash rebinning}(a)).

We note that the spatial resolution does not vary significantly when initially rebinning by a factor of two along each axis. This implies that data recording could be performed using a detector rebin setting, increasing data transfer rates and read time, without compromising resolution.

\subsection{Direct X-ray detectors}
\label{section: photon counting detectors}
\begin{figure*}[hb!]
	\centering
	\begin{tikzpicture}
	\node[anchor=south west] at (0,0) (image1) {\includegraphics[width=.45\textwidth]{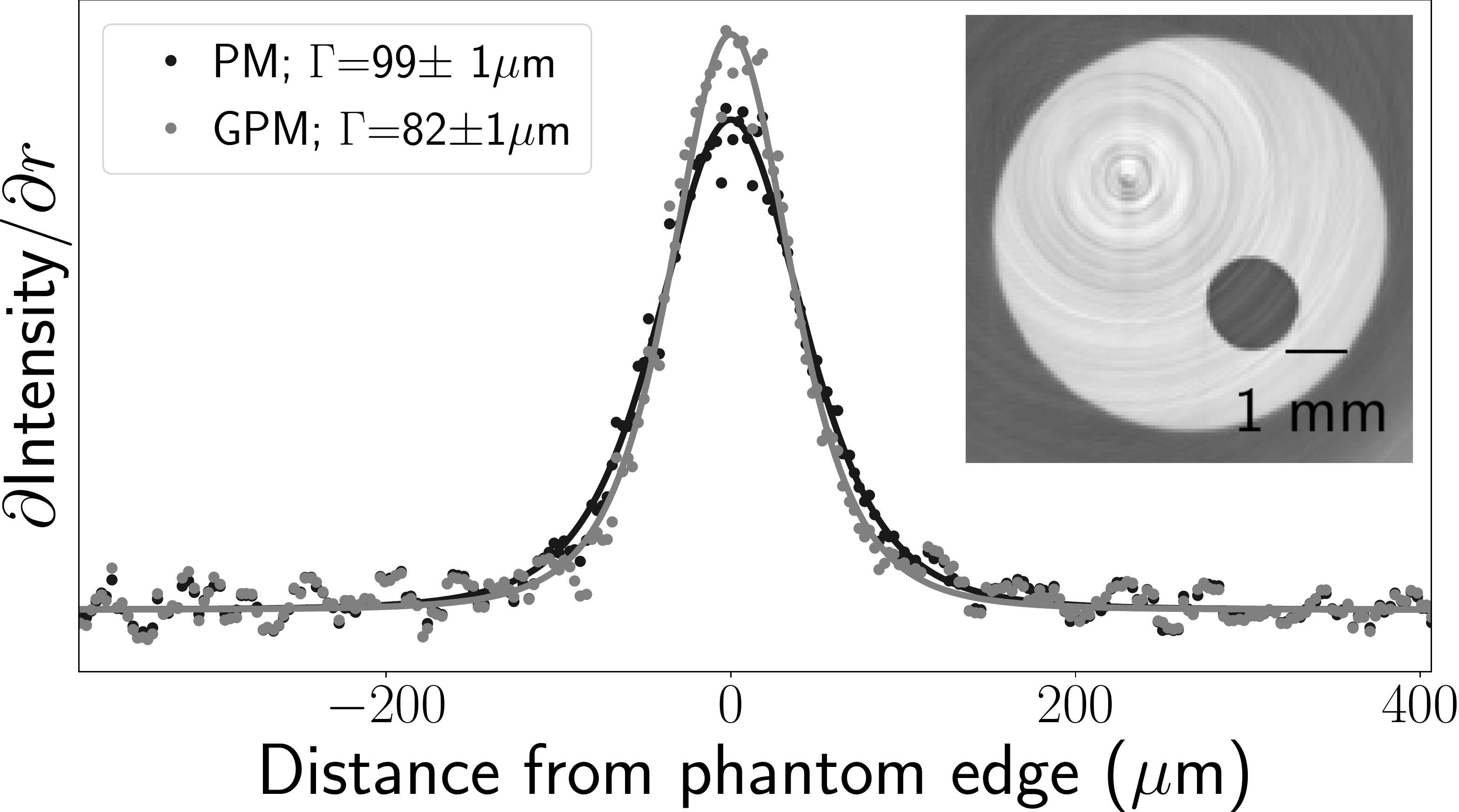}};
	% draw a grid and draw ticks
	\begin{scope}[x={(image1.south east)},y={(image1.north west)}]
		\node[fill=none, font = \large] at (0.03,0.9) {a)};
	\end{scope}
	\end{tikzpicture}\hfill
	\begin{tikzpicture}
	\node[anchor=south west] at (0,0) (image1) {\includegraphics[width=.45\textwidth]{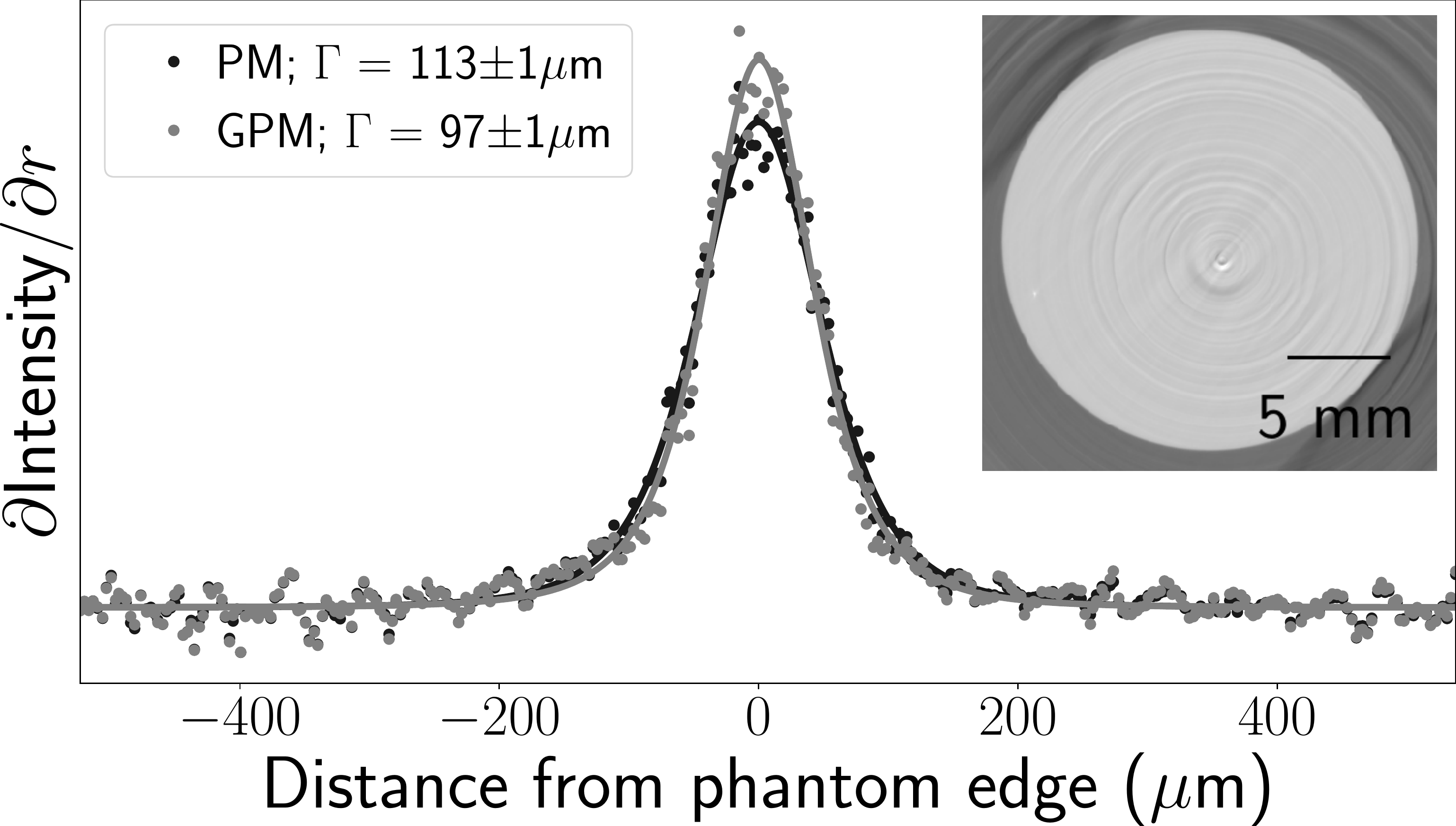}};
	% draw a grid and draw ticks
	\begin{scope}[x={(image1.south east)},y={(image1.north west)}]
		\node[fill=none, font = \large] at (0.03,0.9) {b)};
	\end{scope}
	\end{tikzpicture}
	\captionsetup{font = small}
	\caption{PSFs measured by azimuthally averaging the outer edge PMMA cylinders in CT recorded at \SI{2}{\meter} propagation distance with \SI{24}{\kilo\eV} beam energy using (a) a Advacam Modupix and (b) a LAMBDA photon-counting detector. Voxel size = \SI{55}{\micro\meter}. The PMMA cyclinder used in (a) contains an off-centre circular cavity while the cylinder in (b) is solid PMMA nearing the detector width in size.  }
	\label{figure: widepix analysis}
\end{figure*}
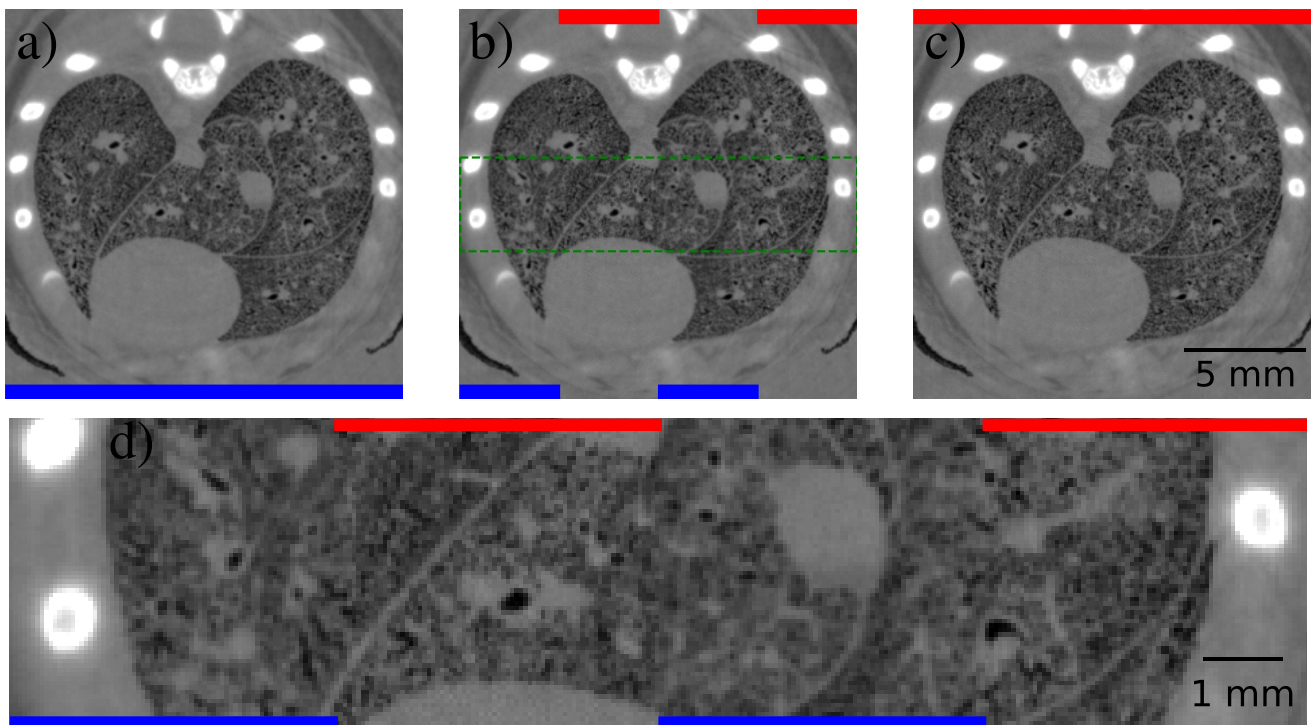
\begin{figure*}[htb!]
	\centering
	\begin{tikzpicture}
	\node[anchor=south west] at (0,0) (image1) {\includegraphics[width=.30\textwidth]{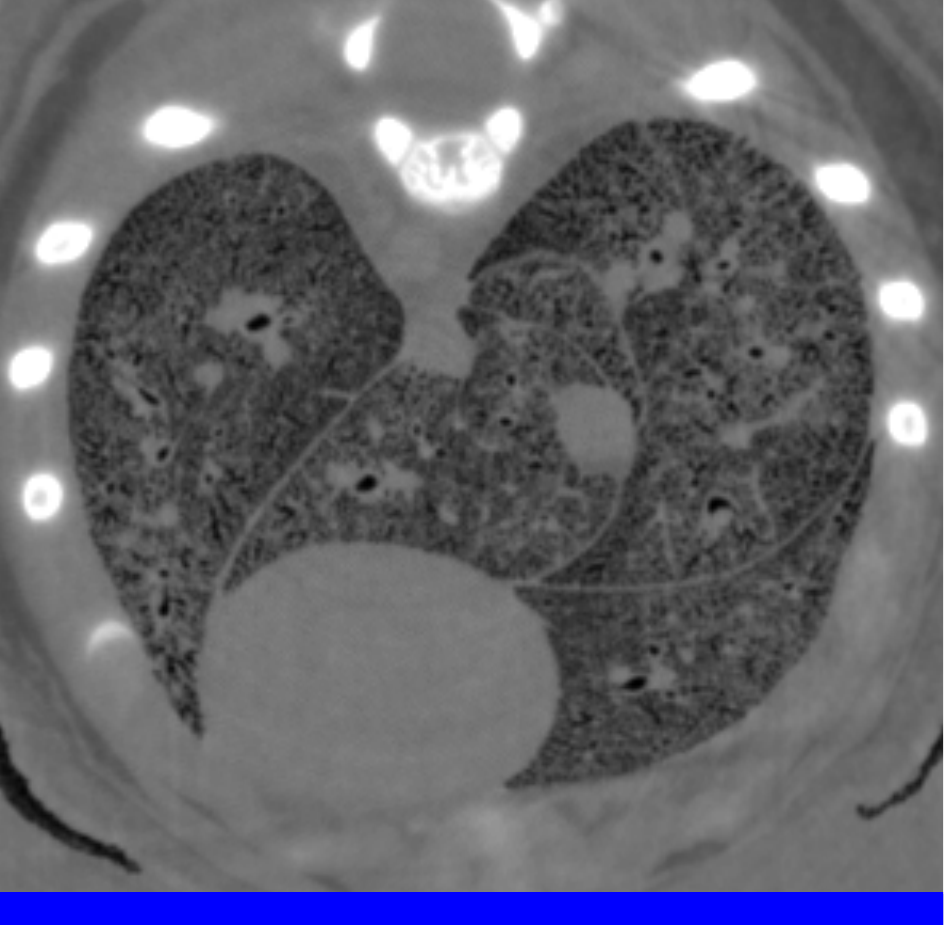}};
	% draw a grid and draw ticks
	\begin{scope}[x={(image1.south east)},y={(image1.north west)}]
		\node[fill=none, font = \huge] at (0.1,0.89) {a)};
	\end{scope}
	\end{tikzpicture}\hfill
	\begin{tikzpicture}
	\node[anchor=south west] at (0,0) (image1) {\includegraphics[width=.30\textwidth]{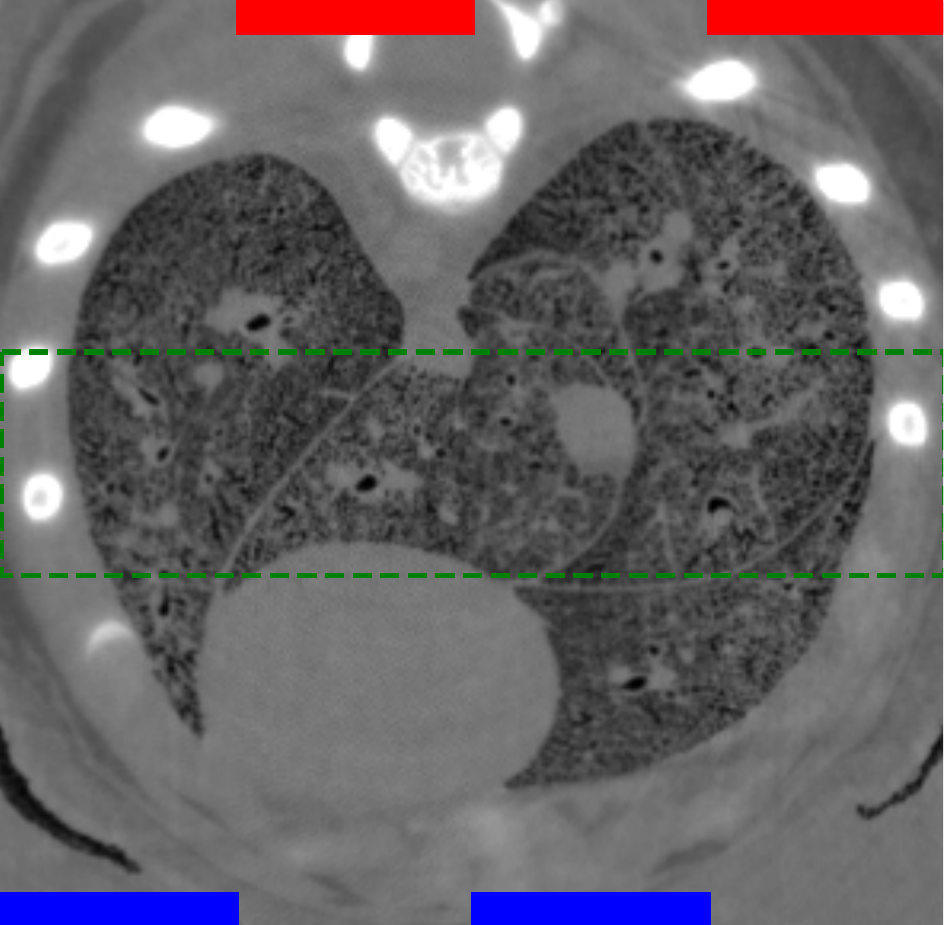}};
	% draw a grid and draw ticks
	\begin{scope}[x={(image1.south east)},y={(image1.north west)}]
		\node[fill=none, font = \huge] at (0.1,0.89) {b)};
	\end{scope}
	\end{tikzpicture}\hfill
	\begin{tikzpicture}
	\node[anchor=south west] at (0,0) (image1) {\includegraphics[width=.30\textwidth]{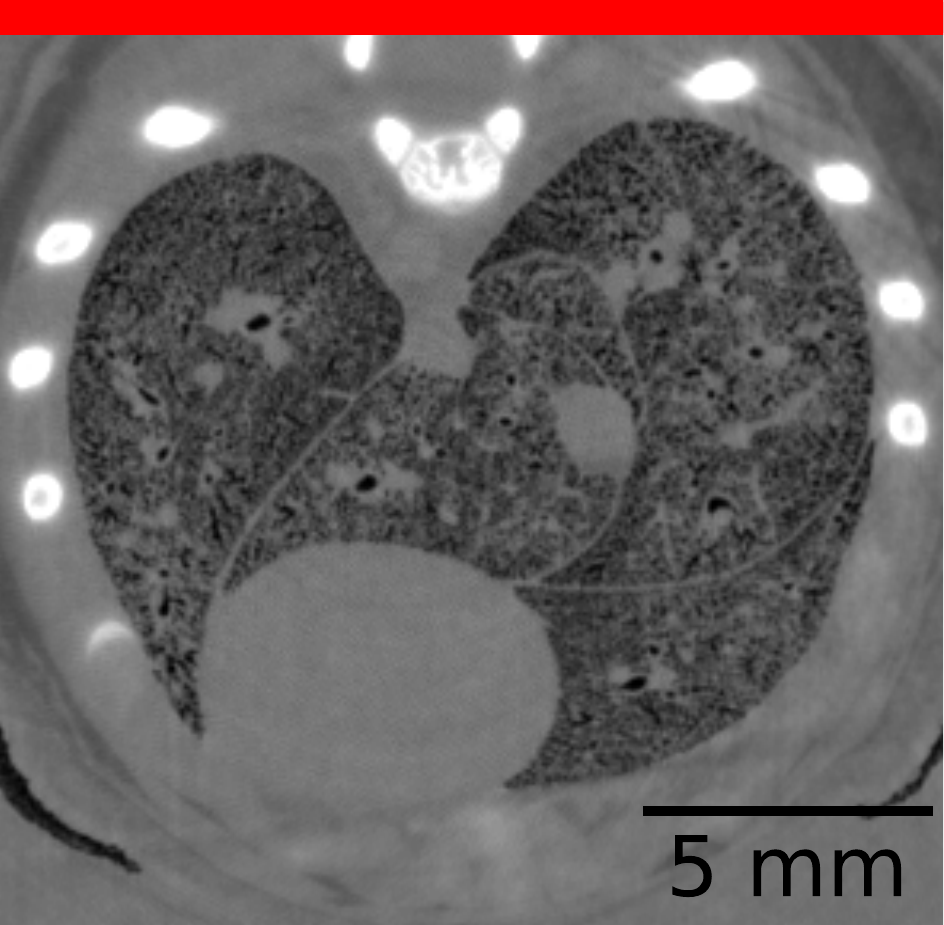}};
	% draw a grid and draw ticks
	\begin{scope}[x={(image1.south east)},y={(image1.north west)}]
		\node[fill=none, font = \huge] at (0.1,0.89) {c)};
	\end{scope}
	\end{tikzpicture}\newline
	\centering
	\begin{tikzpicture}
	\node[anchor=south west] at (0,0) (image1) {\includegraphics[width=.98\textwidth]{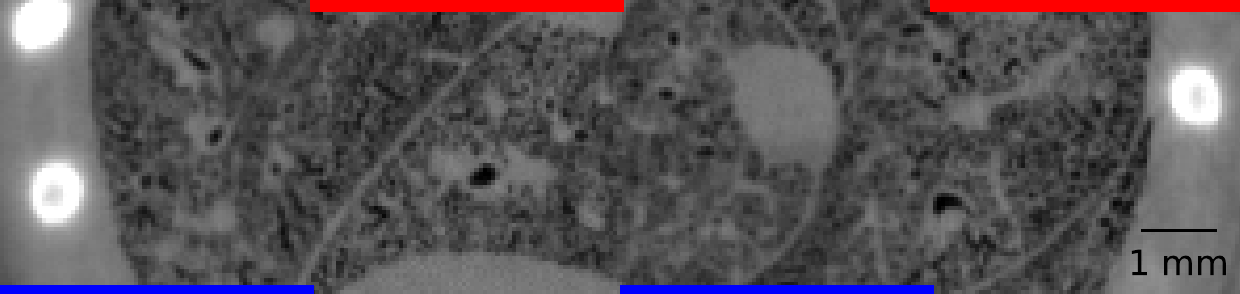}};
	% draw a grid and draw ticks
	\begin{scope}[x={(image1.south east)},y={(image1.north west)}]
		\node[fill=none, font = \huge] at (0.1,0.89) {d)};
	\end{scope}
	\end{tikzpicture}
	\captionsetup{font = small}
	\caption{CT slices of a rat lung reconstructed from phase retrieved projections using either the (a) PM or (c) GPM algorithms. (b) shows an interleaved image of the two methods, with a blue band along the bottom in columns from the PM method and a red band along the top in columns from the GPM, with an inset region in (b) bounded by dashed green lines magnified in (d) for direct comparison of the two methods.}
	\label{figure: rabbit pup lungs}
\end{figure*}
Figure \ref{figure: spatial frequency suppressions}(d) shows that the proportion of high spatial frequencies in the optical system transfer function, which includes phase retrieval, increases as the width of the PSF decreases. From this, we predict that photon counting detectors will provide the best improvement to spatial resolution when using the GPM instead of the PM, since the detector PSF for these systems is inherently limited to the pixel dimensions. To show this, we imaged the same PMMA phantom as was shown in Fig. \ref{figure: little flash rebinning} with an Advacam Modupix photon-counting detector at a \SI{2}{\meter} propagation distance using \SI{24}{\kilo\eV} synchrotron radiation (SPring-8, BL20B2). The detector comprised a Timepix chip with a \SI{55}{\micro\meter} pixel size and silicon sensor, with the threshold set to count X-rays above \SI{12}{\kilo\eV}. Figure \ref{figure: widepix analysis}(a) provides azimuthally averaged PSF fits at the outer boundary of the PMMA phantom using both the PM and GPM algorithms, showing a percentage improvement in spatial resolution of \SI{17\pm2}{\percent} according to equation \eqref{equation: psf percentage improvement} when using the GPM, alongside a drop in SNR from $92\pm14$ to $81\pm16$ ($12\pm35\%$). In a second experiment, we used an X-Spectrum LAMBDA (Large Area Medipix Based Detector Array) 350K photon-counting detector with a \SI{1000}{\micro\meter} thick cadmium telluride (CdTe) scintillator and a \SI{55}{\micro\meter} pixel size to image a larger PMMA phantom. The results are shown in Fig. \ref{figure: widepix analysis}(b) and were recorded at the Imaging and Medical Beamline (IMBL) of the Australian Synchrotron. Although the larger phantom radius was intended to provide more points in the radial averages, we found it was slightly asymmetric and poorly polished, leading us to use only part of the radial arc in our analysis. Defining a radial origin to be from the phantom centre to the top of the image, our radial analysis used an arc from 235-270 degrees, where the shape of the phantom was consistent. These conditions provided a percentage improvement in the PSF FWHM of \SI{14\pm1}{\percent} when using the GPM, accompanied by a drop in SNR from $27\pm6$ to $25\pm5$. 

Next, we show the application of the GPM to a complex sample of biomedical relevance using a direct detector. We collected a CT scan of the thorax of a recently-deceased juvenile rat. The lung inflation was fixed and the rat set in a plastic tube with agarose to prevent motion during imaging. Imaging was performed, again using the LAMBDA photon counting detector, at a \SI{2}{\meter} propagation distance and a beam energy of \SI{26}{\kilo\eV} at the IMBL. The energy threshold of the detector was set to \SI{6}{\kilo\eV} and phase retrieval was performed on the lung-air interface, using water as an analogous material. Figure \ref{figure: rabbit pup lungs}(b) shows an interleaved CT slice where the red pixel columns are from a GPM-reconstructed slice, Fig. \ref{figure: rabbit pup lungs}(c), and the blue columns are from a PM-reconstructed slice, Fig. \ref{figure: rabbit pup lungs}(a). At small scales, the difference between each resolution is subtle; however, expanding the inset in Fig. \ref{figure: rabbit pup lungs}(c) as in Fig. \ref{figure: rabbit pup lungs}(d), shows a much clearer difference between the two methods. Where the PM columns look blurred, the GPM columns appear sharper, with higher definition between the airways (including the alveoli) and the soft tissues. We reiterate that this straightforward improvement to spatial resolution achievable by the GPM when using direct detectors is a consequence of the single-pixel PSFs they possess and hence would equivalently improve data recorded with indirect detectors of similar PSF. Similarly, methods allowing sub-pixel resolution through localisation of charge clouds caused by X-ray interaction with the detector, performed with indirect \cite{lumb_event_1988,oconnell_photon-counting_2020,nowak_sub-pixel_2015} and direct detectors\cite{cartier_micrometer-resolution_2016,dreier_tracking_2020}, would likely benefit further from the GPM rederivation. 

\section{Rederivation of additional phase retrieval algorithms using discrete mathematics}
\label{section: established algorithms}
The GPM rederivation is achieved through a modified version of the Fourier derivative theorem. In a continuous domain, the Fourier derivative theorem is expressed as
\begin{align}
\nabla \mathcal{F}[f(x,y)] &= - i \textbf{k}_{\perp \text{PM}} \mathcal{F}[f(x,y)]\label{equation: fourier derivative theorem},
\end{align}
but in a discrete domain we need to incorporate the discrete representation of the Laplacian, described in  \cite{paganin_boosting_2020} through the 5 point approximation, giving 
\begin{align}
\nabla \text{DFT}[f(x,y)] &= - i \textbf{k}_{\perp \text{GPM}} \text{DFT}[f(x,y)] \label{equation: discrete fourier derivative theorem}.
\end{align}
Given the improvement to spatial resolution comes from the modified Fourier derivative theorem, we infer the same benefit can be replicated in other phase retrieval algorithms that use Fourier transforms. Here, we consider two algorithms used commonly in propagation-based phase contrast imaging that also utilize the discrete Fourier transform. First, we look at the two-material phase retrieval algorithm of Beltran \textit{et al.} \cite{beltran_2d_2010}, followed by the dual-energy algorithm for decomposing images into their photoelectric and Compton scatter components \cite{gursoy_single-step_2013}. We provide experimental validation of improvements in spatial resolution, achieved through implementation of the discrete Fourier derivative theorem in equation \eqref{equation: discrete fourier derivative theorem}. We use these to demonstrate the benefit of applying equation \eqref{equation: discrete fourier derivative theorem} to derivations in place of equation \eqref{equation: fourier derivative theorem}. Note that this is not an exhaustive list of all the possible extensions, and we welcome further applications of this approach. 

\subsection{Two-material phase retrieval algorithm}
\label{section: established algorithms - beltran}
\begin{figure}[tb!]
	\centering
	\includegraphics[width = 0.45\textwidth]{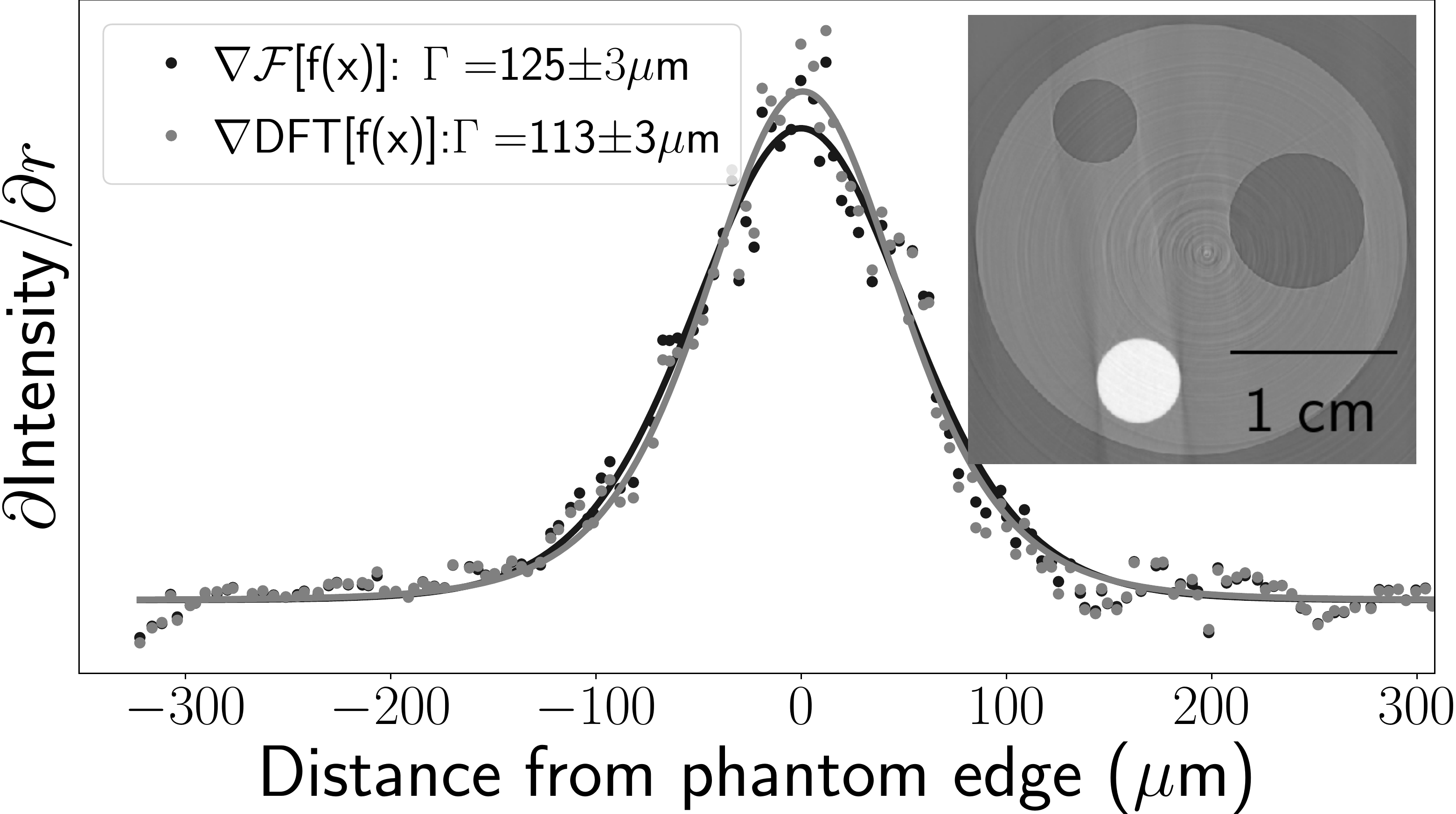}
	\captionsetup{font = small}
	\caption{Pearson VII fit comparison of spatial resolution achieved through a PM and GPM application of the Beltran two-material phase retrieval algorithm, featuring a PMMA phantom with aluminium inset. Data was recorded using the LAMBDA detector at a propagation distance of \SI{2}{\meter}, and spatial resolution analysis was performed using the aluminium-PMMA material interface.}
	\label{figure: beltran comparison}
\end{figure}

We focus on the highly-stable two-material propagation-based phase retrieval algorithm developed by Beltran \textit{et al.} \cite{beltran_2d_2010}. This algorithm was designed used to correctly reconstruct the projected thickness of one material embedded within another, rather than just focussing on a single material within air. This method requires knowledge of the complex refractive indices of both materials and, for quantitative measures, their total projected thickness
\begin{align}
A(x,y) &= T_1(x,y) + T_2(x,y),
\end{align}
with $T_1$ and $T_2$ representing the spatially variant thickness of each material. However, when combined with CT, it is not necessary to isolate materials in projection as they are spatially separated by backprojection. Croton \textit{et al.} \cite{croton_situ_2018} showed that it is therefore not necessary to know $A(x,y)$ for use with CT. This simplifies the algorithm to a low-pass filter that specifically matches the fringe enhancement effects at the boundary between the two materials, creating a noise-reduced representation of the absorption contrast image, given by
\begin{align}
I_{\text{B-PM}}(x, y, z = 0) &= \mathcal{F}^{-1}\left[\frac{\mathcal{F}[I(x, y, z = \Delta)/I_0]}{1 + \frac{(\delta_2 - \delta_1) \Delta}{\mu_2 - \mu_1}\textbf{k}_{\perp \text{PM}}^2}\right] \label{equation: beltran paganin method thickness}.
\end{align}
Modifying equation \eqref{equation: beltran paganin method thickness} to use a discrete representation of the Fourier derivative theorem, equation \eqref{equation: discrete fourier derivative theorem}, gives
\begin{align}
I_{\text{B-GPM}}(x, y, z = 0) &= \text{DFT}^{-1}\left[\frac{\text{DFT}[I(x, y, z = \Delta)/I_0]}{1 + \frac{(\delta_2 - \delta_1) \Delta}{\mu_2 - \mu_1}\textbf{k}_{\perp \text{GPM}}^2}\right]. \label{equation: beltran general paganin method thickness}
\end{align}
We imaged a cylindrical PMMA phantom with three cylindrical cavities; one filled with an aluminium rod and the other two with air. Figure \ref{figure: beltran comparison} shows a radial PSF fit around the Aluminium inset, recorded using the LAMBDA photon-counting detector at a propagation distance of \SI{2}{\meter} and an X-ray energy of \SI{40}{\kilo\eV} at the IMBL. Here the GPM rederivation shows an \SI{11\pm3}{\percent} improvement in spatial resolution at the aluminium-PMMA boundary compared to the original method. 

\subsection{Dual-energy material decomposition}
\label{section: established algorithms - Florian}
\begin{figure*}[b!]
	\centering
	\begin{tikzpicture}
	\node[anchor=south west] at (0,0) (image1) {\includegraphics[width=.45\textwidth]{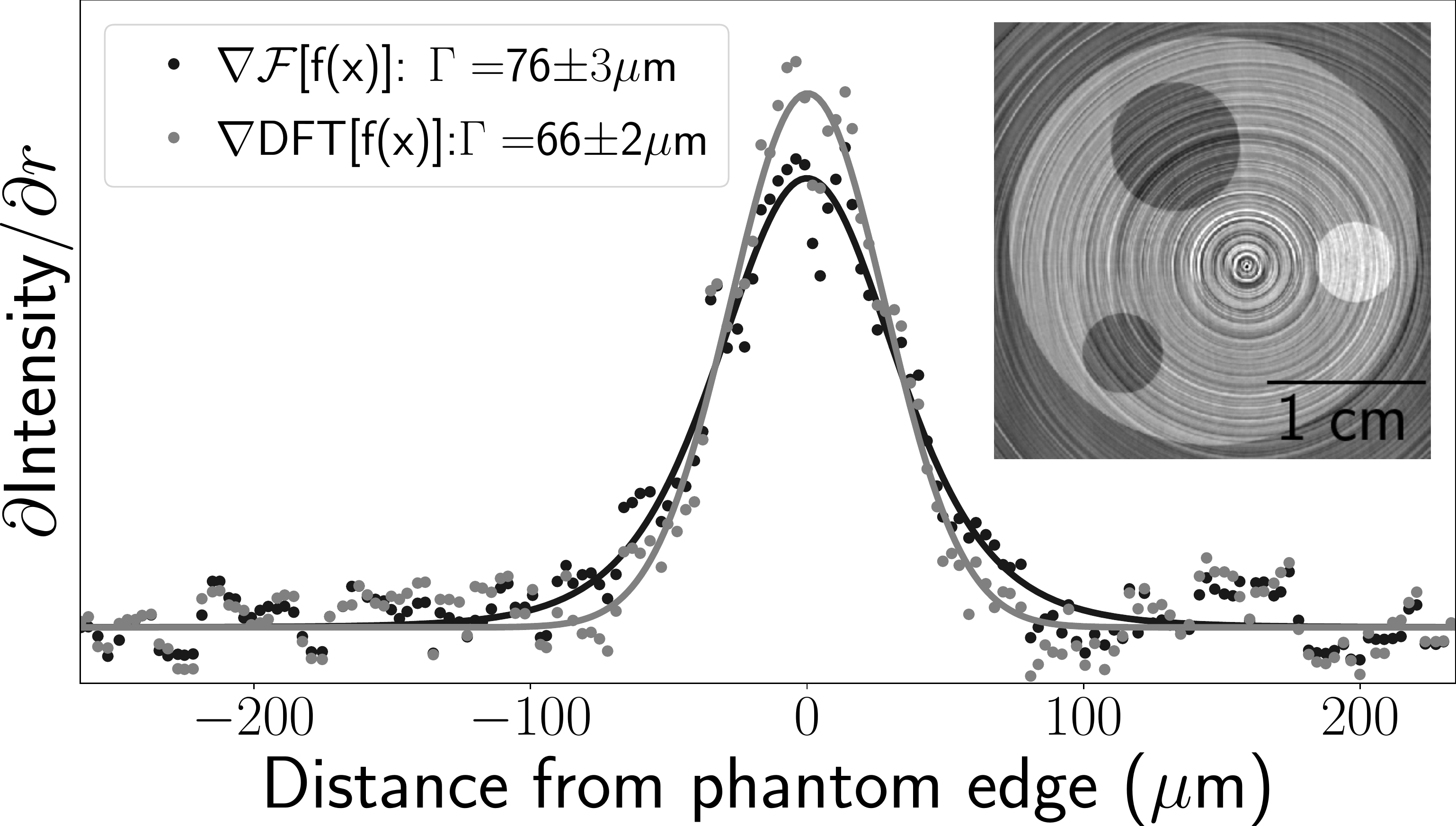}};
	% draw a grid and draw ticks
	\begin{scope}[x={(image1.south east)},y={(image1.north west)}]
		\node[fill=none, font = \large] at (0.03,0.9) {a)};
	\end{scope}
	\end{tikzpicture}\hfil
	\begin{tikzpicture}
	\node[anchor=south west] at (0,0) (image1) {\includegraphics[width=.45\textwidth]{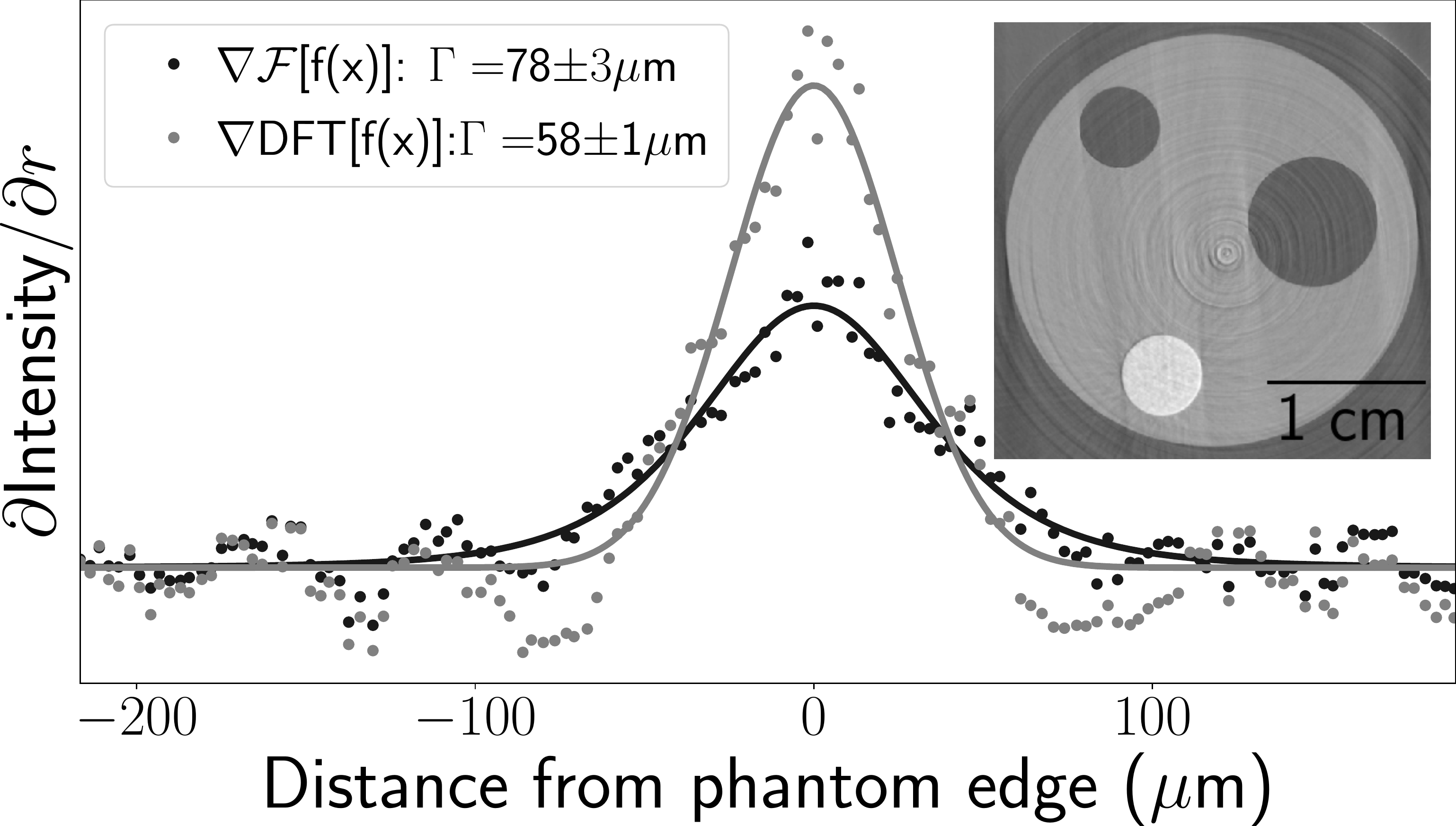}};
	% draw a grid and draw ticks
	\begin{scope}[x={(image1.south east)},y={(image1.north west)}]
		\node[fill=none, font = \large] at (0.03,0.9) {b)};
	\end{scope}
	\end{tikzpicture}
	\captionsetup{font = small}
	\caption{PSF comparison of the dual-energy phase retrieval images showing reconstructed slices of electron density maps shown in the graph inset. PSFs have been fit with Pearson VII functions for quantifying the PSF FWHM. (a) shows a spatial resolution improvement of \SI{16\pm6}{\percent} at \SI{1}{\meter} propagation distance, while (b) shows the same phantom at \SI{2}{\meter} propagation distance showing a \SI{29\pm2}{\percent} improvement, likely due to the higher SNR afforded by low-pass filtering in the phase retrieval algorithm.}
	\label{figure: Florian AM decomposition}
\end{figure*}%
Phase retrieval algorithms for PBI have recently been developed to use the phase and attenuation components to enable material decomposition \cite{schaff_material_2020}\cite{li_Quantitative_2020} or to separate attenuation from phase effects \cite{gursoy_single-step_2013}. The methods rely on the acquisition of PBI images at a minimum of two different X-ray wavelengths. Here we further generalise the phase retrieval algorithm of Gursoy and Das \cite{gursoy_single-step_2013}, which uses the Alvarez-Macovski (AM) model of X-ray attenuation \cite{alvarez_energy-selective_1976},  and begins from the linearised TIE 
\begin{align}
-\mathcal{F}\ln I(E) &= \mathcal{F}\left[\int \mu (E)dz\right] + \Delta k_{\perp \text{PM}}^2 \mathcal{F}\left[\int \delta(E)dz\right],
\end{align}
where $I(E)$, $\mu(E)$ and $\delta(E)$ are discrete image maps of the spatially variant contact intensity, linear attenuation coefficient, and refractive index decrement and are all functions of energy, $E$. Integrations are performed along the optical axis, labelled as z. To change basis, $\mu(E)$ is represented as a linear combination of photoelectric absorption and Compton scattering. Photoelectric absorption is proportional to $E^{-3}$ while Compton scattering is proportional to both the electron density, $\rho_e$, and the Klein-Nishina cross section, $\sigma_{\mathrm{KN}}(E)$. Since $\delta(E)$ is also a function of electron density $\delta(E)=(\rho_e h^2 c^2 r_e)/(2\pi E^2)$, where $h$ is Planck's constant, $c$ is the speed of light and $r_e$ is the classical electron radius, allows the Compton and phase terms to be coupled together. This coupling allows us to solve for the attenuation and phase, or alternatively the projected electron density, using PBI images acquired using at least two energies as a matrix equation:
\begin{align}
\begin{bmatrix} E_A^{-3} & \sigma_{KN}(E_A) + \chi_A k_{\perp \text{PM}}^2 \\  E_B^{-3} & \sigma_{KN}(E_B) + \chi_B k_{\perp \text{PM}}^2 \end{bmatrix} \begin{bmatrix} \mathcal{F}P \\ \mathcal{F}\int\rho_e \end{bmatrix} &= \begin{bmatrix} -\mathcal{F}\ln I(E_A) \\ -\mathcal{F}\ln I(E_B) \end{bmatrix}, \label{eq: matrix equation} 
\end{align} 
where $\chi_A = (h^2 c^2 r_e \Delta) / (2\pi E_A^2)$ and $\chi_B = (h^2 c^2 r_e \Delta)(2\pi E_B^2)$. Equation \eqref{eq: matrix equation} can then be inverted to solve for P, the photoelectric component, and $\int\rho_e$. Details of this inversion can be found in Schaff $et al.$ \cite{schaff_spectral_2020}. Schaff \textit{et al.} showed that the reconstruction of $\int\rho_e$ is highly robust as the combined phase and Compton signals provide a solution that includes a low-pass Fourier filter term that smooths noise. Such smoothing is akin to the single image phase retrieval algorithms in Paganin \textit{et al.} \cite{paganin_simultaneous_2002} and Beltran \textit{et al.} \cite{beltran_2d_2010}. Here, we generalise equation \eqref{eq: matrix equation} using a discretised version of the Fourier derivative theorem by replacing $k_{\perp \text{PM}}^2$ with $k_{\perp \text{GPM}}^2$. We focus our study on the stable electron density results.
\\ \\
To explore the benefit of the modified derivation in improving spatial resolution, we performed 180$^{\circ}$ CTs of a dual-material phantom, again using the LAMBDA detector at IMBL. We used the same cylindrical PMMA and aluminium phantom described in the previous section. Given that the highly-attenuating aluminium rod was approximately \SI{5}{\mm} in diameter, we opted for higher energies, \SI{30}{\kilo\eV} and \SI{40}{\kilo\eV}, than in our previous studies, to reduce beam hardening artefacts. Figure \ref{figure: Florian AM decomposition} shows slices of the electron density maps calculated from equation \eqref{eq: matrix equation} using dual-energy recordings at \SI{30}{\kilo\eV} and \SI{40}{\kilo\eV}. Figure \ref{figure: Florian AM decomposition}(a) uses a propagation distance of \SI{1}{\meter} and \ref{figure: Florian AM decomposition}(b) uses \SI{2}{\meter} propagation distance. The 1 m propagation distance provides a region entirely within the validity of the TIE, but possesses significant ring artefacts due to limited SNR enhancement from phase retrieval. The \SI{2}{\meter} distance provides a stronger SNR enhancement, while sitting at the edge of the TIE validity; signified by the slight fringes on either side of the Pearson VII fits in \ref{figure: Florian AM decomposition}(b). At a \SI{1}{\meter} propagation distance, equation \eqref{equation: psf percentage improvement} gives a \SI{16\pm6}{\percent} improvement in the PSF, while the \SI{2}{\meter} propagation gives a \SI{29\pm2}{\percent} improvement. Each PSF fit uses a \SI{30}{\degree} arc of the outer PMMA edge, (a) using the angular range \SIrange{202.5}{247.5}{\degree} and (b) using the range \SIrange{45}{90}{\degree}. Overall, we find using the discretized Fourier derivative theorem provides increased spatial resolution when applied to the Alvarez-Macovski-based dual-energy reconstruction of electron density and would likely improve resolution for material decomposition too. 
% Given the significant ring artefacts in figure \ref{figure: Florian AM decomposition}(a) we attempted to use a ring correction method from Vo \textit{et al.} \cite{vo_superior_2018} which reorders the sinogram with intensity along the radial direction and then applies a median filter before restoring the original order. Although the method very successfully removed the ring artefacts in figure \ref{figure: Florian AM decomposition}(a) we found it unpredictable as to whether it improved spatial resolution or worsened it.

\section{Conclusion}
Paganin \textit{et al.} \cite{paganin_boosting_2020} presented a rederivation of the PBI phase retrieval algorithm of Paganin \textit{et al.} \cite{paganin_simultaneous_2002}, implementing a discretized form of the Laplacian, and provided a theoretical grounding for its ability to increase spatial resolution in processed images. We expand on this result by first considering how the point spread function affects the algorithm. Using theory and simulation, we demonstrated that for detectors with a PSF larger than the pixel size, the spatial frequency filter used in discretized phase retrieval becomes comparable to the standard approach, hence the total gain in spatial resolution may be minimal. We verified experimentally that the magnitude of any improvement is highly dependent upon the width of the detector PSF using PBI-CT scans with both direct and indirect detectors. For detector systems with PSF widths of several pixels or more, typical of unbinned indirect detectors, we saw negligible improvement to spatial resolution due to the broad PSF suppressing high spatial frequency content before it could be boosted; however, by rebinning the same data to reduce the effective PSF width (expressed in pixels), the GPM achieved up to a \SI{16\pm2}{\percent} improvement in spatial resolution compared to PM, only after binning by a factor of 8. Conversely, detector systems effectively possessing single pixel width PSFs, commonly found in indirect detectors, showed up to \SI{17\pm2}{\percent} improvement in spatial resolution.. We next extended the concept of the more accurate discrete Fourier formalism to other stable phase retrieval algorithms. For the two material algorithm of Beltran \textit{et al.} \cite{beltran_interface-specific_2011}, we found the spatial resolution was improved by \SI{11\pm3}{\percent} using a direct detector. Furthermore, 3D maps electron density recovered using the dual-energy algorithm of Schaff \textit{et al.} \cite{schaff_spectral_2020}  with \SI{29\pm2}{\percent} resolution improvement. We speculate that similar spatial resolution improvements could be gained for other algorithms that use the Fourier derivative theorem in a discrete setting. 

Although the improvements in resolution afforded by the discrete phase retrieval algorithms may only be modest, the most important benefit is the knowledge that the object has been reconstructed more faithfully than was previously achieved. Given the increasing popularity of direct, photon-counting detectors for dose reduction and spectral imaging, and the superior performance of these discrete Fourier transform-based algorithms when using photon-counting detectors, we anticipate that these rederived phase retrieval algorithms will be of increasing importance in future phase contrast work. 

\section*{Data availability}
The datasets used and/or analysed for this manuscript are available from the corresponding author on reasonable request.

\bibliography{Pollock_Bibfile}
\section*{Acknowledgements}
We would like to thank David Paganin for helpful discussion on the GPM method and, likewise, Florian Schaff for discussion on implementation of the dual-energy phase retrieval algorithm.
This work was funded in part by the Japan Synchrotron Radiation Research Institute (JASRI) under Project 2019B0150, in part by the Future Fellowship Schemes under Grant FT160100454 and Grant FT180100374, in part by the International Synchrotron Access Program (ISAP) managed by the Australian Synchrotron, a part of the Australian Nuclear Science and Technology Organisation (ANSTO) under Grant AS/IA193/16034, in part by the Australian Synchrotron under Grant AS193/IMBL/15223. We also thank C.J. Hall and A. Maksimenko for aiding in experimentation at the IMBL. The work of James A. Pollock was supported in part by the Research Training Program (RTP) Scholarship and in part by the J. L. Williams Top Up Scholarship. 

\section*{Author contributions statement}
J.A. Pollock was a part of all data collection and performed the subsequent anaylsis and manuscript writing. M.K., K.M., L.C.P.C., M.K.C and G.R. contributed to collecting data and editing the manuscript. N.Y and H.S as beamline scientists at the Japanese synchrotron Spring-8 had a crucial role in the experiments performed on the 20B2 beamline. All authors contributed to the manuscript. 

\end{document}